\documentclass[prd,aps,nofootinbib,nolongbibliography,superscriptaddress]{revtex4-2}
\usepackage{epsfig}
\usepackage{amsmath, slashed}
\usepackage{xcolor}
\usepackage{appendix}
\usepackage{dsfont}

\usepackage[normalem]{ulem}

\usepackage{hyperref}

\usepackage{bm}
\usepackage{amsfonts}
\allowdisplaybreaks
\begin{document}

\title{Timelike Compton scattering on a deuteron target }

\author{Jing Han}
\affiliation{School of Physics, Zhengzhou University, Zhengzhou, Henan 450001, China}
\author{ Bo-Qiang Ma}
\email{ mabq@pku.edu.cn}
\affiliation{School of Physics, Zhengzhou University, Zhengzhou,  Henan 450001, China}
\author{Bernard Pire }
\email{bernard.pire@polytechnique.edu}
\affiliation{CPHT, CNRS, \'Ecole polytechnique, Institut Polytechnique de Paris, Palaiseau, France}
\author{Qin-Tao Song}
\email{songqintao@zzu.edu.cn}
\affiliation{School of Physics, Zhengzhou University, Zhengzhou, Henan 450001, China}

\begin{abstract}
{We investigate timelike Compton scattering (TCS) on a deuteron target as a probe of the three-dimensional partonic structure of the deuteron encoded in generalized parton distributions (GPDs). Such measurements may provide new insight into the internal structure of the deuteron,  and help to clarify the unresolved tensor-polarization puzzle. We present numerical estimates of the leading-order  (in the strong coupling $\alpha_s$) TCS cross section in the kinematic region relevant to future measurements at Jefferson Lab. The deuteron GPDs are evaluated within a convolution model, while the Bethe-Heitler contribution is calculated using a phenomenological parametrization of the deuteron electromagnetic form factors. In addition, we derive the theoretical expressions for the next-to-leading-order corrections to deuteron TCS, with particular emphasis on contributions from the gluon transversity GPDs. We show that these contributions can be isolated through their characteristic angular dependence in the center-of-mass frame of the produced lepton pair. Our results demonstrate the potential of deuteron TCS as a complementary probe of both the quark and gluon structure of the deuteron at Jefferson Lab.}
\end{abstract}

\date{\today}

\maketitle


\section{Introduction}

The deuteron, as the simplest bound nuclear system and a spin-1 hadron, provides a unique laboratory for investigating the interplay between nucleonic and partonic degrees of freedom in nuclei. At the level of one-dimensional parton distribution functions (PDFs), its spin-1 nature gives rise to tensor-polarized distributions that have no counterpart in spin-$1/2$ hadrons~\cite{Hoodbhoy:1988am,Close:1990zw,Kumano:2010vz,Kumano:2016ude}. In 2005, the HERMES Collaboration reported the first measurement of the leading-twist tensor structure function $b_1(x)$~\cite{HERMES:2005pon}, which is directly related to the tensor-polarized PDF $f_{1\scriptscriptstyle LL}(x)$. The measured $b_1$ was found to be substantially larger than predictions from conventional calculations~\cite{Khan:1991qk,Cosyn:2017fbo}, in which the deuteron is described predominantly as a weakly bound proton-neutron system. This discrepancy has motivated proposals involving non-nucleonic degrees of freedom in the deuteron~\cite{Miller:2013hla}, suggesting that its partonic structure may not be fully captured by a purely nucleonic description. The origin of the unexpectedly large tensor-polarized structure is often referred to as the tensor-polarization puzzle of the deuteron. To clarify this issue, precision measurements of tensor-polarized PDFs are planned or proposed at several facilities~\cite{Farooq:2026kky,Keller:2022abm,Clement:2023eun,Poudel:2025nof,Arbuzov:2020cqg}. In parallel, theoretical studies of tensor-polarized partonic structure have recently been extended beyond the leading-twist approximation, including investigations of higher-twist PDFs and corrections in spin-1 hadrons~\cite{Koile:2011aa,Detmold:2005iz,Kumano:2021fem,Kumano:2021xau,Qiao:2024bgg,Cosyn:2024drt,Kumano:2026xxv,Pascalutsa:2026ubk}.

In addition to one-dimensional PDFs, multidimensional partonic distributions can provide substantially richer information on the internal structure of the deuteron. Transverse-momentum-dependent parton distributions (TMDs) encode correlations between the longitudinal and transverse motion of partons~\cite{Bacchetta:2000jk,Bacchetta:2001rb,Kumano:2020ijt,Ninomiya:2017ggn,Cotogno:2017puy,Zhao:2025vol}, while generalized parton distributions (GPDs) provide access to correlations between the longitudinal momentum and transverse spatial distributions of partons~\cite{Muller:1994ses,Ji:1996ek,Radyushkin:1996nd,Diehl:2003ny,Belitsky:2005qn,Guo:2026swz,Berger:2001zb,Dong:2013rk,Cosyn:2018rdm,Cosyn:2018thq}. Moreover, the second moments of hadronic GPDs are related to gravitational form factors~\cite{Polyakov:2018zvc,Kumano:2017lhr,Taneja:2011sy,Burkert:2023wzr,Freese:2021qtb,Lorce:2025ayr,Fujii:2025pkv,Polyakov:2019lbq,Cosyn:2019aio}, which encode fundamental information on the energy, momentum, angular momentum, and mechanical properties of the deuteron. Hard exclusive reactions provide experimental access to GPDs within the framework of collinear QCD factorization. Both deeply virtual Compton scattering (DVCS)~\cite{Benali:2020vma,Cano:2002ph,Kirchner:2002zu,Kirchner:2003wt,Anikin:2010ng,Anikin:2011aa,Cano:2003ju} and deeply virtual meson production (DVMP)~\cite{Cano:2003ju,JeffersonLabHallA:2017hky} have been investigated theoretically and experimentally as probes of deuteron GPDs. Timelike Compton scattering (TCS)~\cite{Berger:2001xd} is the timelike counterpart of DVCS,  and provides complementary sensitivity to Compton form factors (CFFs) through the interference of TCS and Bethe-Heitler (BH) processes and lepton angular dependence. This complementarity has been extensively discussed in \cite{Mueller:2012sma,Grocholski:2019pqj}.
 However, studies of TCS on the deuteron remain comparatively limited. By contrast, the first measurement of TCS on the proton has already been reported by the CLAS Collaboration at Jefferson Lab~\cite{CLAS:2021lky} and near future studies are planned~\cite{Alvarado:2026ggy}. Extending TCS measurements to a deuteron target would therefore provide a complementary opportunity to explore the considerably richer GPD structure associated with a spin-1 hadron. We restrict ourselves to the coherent channel $\gamma d \to d\,\ell^+\ell^-$. The competing breakup reaction $\gamma d \to pn\,\ell^+\ell^-$, including $NN$ final-state interactions and the collinear singularities in the lepton azimuthal angle, has been calculated in Ref.~\cite{Cai:2022qza}; its experimental subtraction is a prerequisite for isolating the coherent TCS signal discussed below.

In this work, we investigate coherent TCS on a deuteron target
in the kinematic region relevant to future measurements at Jefferson Lab. We first present the formalism for the TCS and  BH amplitudes for a spin-1 target and derive the corresponding differential cross sections. Numerical estimates at leading order (LO) and leading twist \footnote{Estimates for kinematical higher twist effects~\cite{Braun:2011dg,Braun:2011zr} have been discussed for TCS on a spin zero target~\cite{Martinez-Fernandez:2025gub}. The twist three effects of both  kinematical and dynamical sources on the deuteron target are also considered in Ref.~\cite{Anikin:2010ng}. These effects are not much greater on heavy targets, as discussed for DVCS on the helium-4 nucleus in \cite{Martinez-Fernandez:2026web}. We estimate these effects to be lower than 10 per cent in the TCS cross section at $Q^2>5$ GeV$^2$ in the low $-t$ domain discussed here, and we thus do not consider them further. } are obtained using a convolution model for the deuteron GPDs,  together with a phenomenological parametrization of the deuteron electromagnetic form factors (FFs) for the BH contribution. We examine the relative magnitude of the TCS, BH, and interference contributions and discuss their implications for future experimental measurements. We further present the  next-to-leading-order (NLO) formalism for the deuteron TCS cross sections, with particular attention to the gluon transversity GPDs, and discuss how their characteristic contribution may be isolated through the angular distribution of the produced lepton pair. Our study provides a theoretical basis for using deuteron TCS to explore both the quark and gluon three-dimensional structure of the deuteron.

\section{generalized parton distributions in TCS}

We now extend the analysis to a deuteron target and consider the exclusive photoproduction of a timelike lepton pair,
\begin{equation}
\gamma(q_{1})+d(p_{1})
\to \gamma^{\ast}(q_{2})+d(p_{2})
\to l^{-}(k_{1}) l^{+}(k_{2}) d(p_{2}).
\end{equation}
The relevant kinematic variables are defined as
\begin{equation}
Q^{2} = (q_{2})^{2}>0, \qquad
s=(p_{1}+q_{1})^{2}, \qquad
t=(p_{2}-p_{1})^{2}=\Delta^{2},\qquad
(p_{1})^{2}=(p_{2})^{2}=m^{2}, \qquad
P=2\bar{P}=p_{1}+p_{2},
\end{equation}
where $m$ denotes the deuteron mass.

Following Ref.~\cite{Moutarde:2013qs}, we introduce the
light-cone vectors
\begin{equation}
n=cq_{1}, \qquad
\tilde{n}
=\frac{1}{c(q_{1}\cdot p_{1})}
\left[p_{1}
-\frac{m^{2}}{2(q_{1}\cdot p_{1})}q_{1}
\right],
\label{eqn}
\end{equation}
which satisfy
\begin{equation}
n^{2}=\tilde{n}^{2}=0, \qquad
n\cdot\tilde{n}=1.
\end{equation}
Here, $c$ is an arbitrary nonzero normalization constant, which we
choose in this work as
\begin{equation}
c=\frac{2}{(p_{1}+p_{2})\cdot q_{1}}.
\end{equation}
For a generic
four-vector $a^\mu$, the light-cone components are defined as
\begin{equation}
a^{+}=a\cdot n, \qquad
a^{-}=a\cdot\tilde{n}.
\end{equation}
The corresponding transverse tensors are
\begin{equation}
\begin{aligned}
g_{T}^{\mu\nu}
&=
g^{\mu\nu}
-\frac{1}{q_{1}\cdot p_{1}}
\left[
p_{1}^{\mu}q_{1}^{\nu}
+p_{1}^{\nu}q_{1}^{\mu} -\frac{m^{2}}{q_{1}\cdot p_{1}}q_{1}^{\mu}q_{1}^{\nu}
\right],\
\varepsilon_{T}^{\mu\nu}
&=
\frac{\varepsilon^{\mu\nu p_{1}q_{1}}}
{q_{1}\cdot p_{1}},
\end{aligned}
\end{equation}
where we adopt the convention $\varepsilon^{0123}=1$.

\begin{figure}[htbp]
\centering
\includegraphics[width=0.8\textwidth]{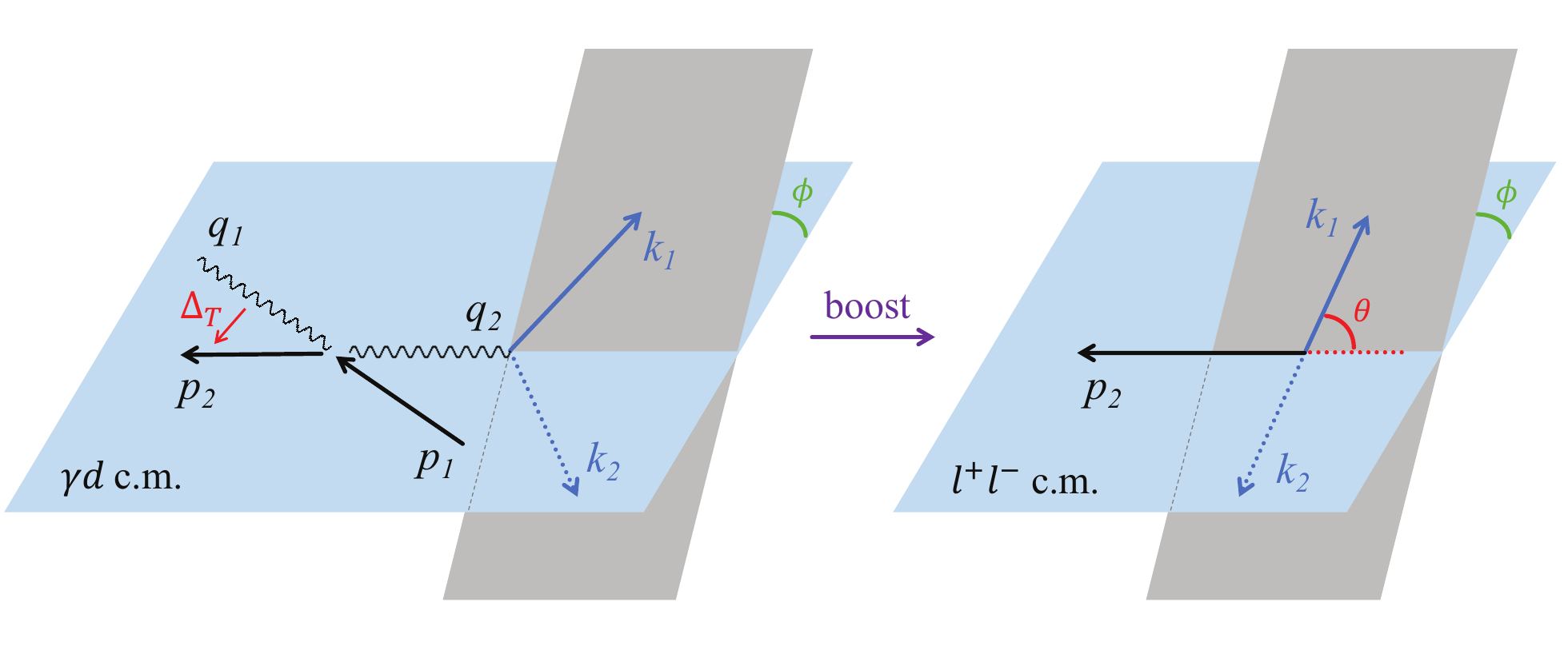}
\caption{Kinematics of deuteron TCS in the $\gamma d$ and dilepton center-of-mass frames}
\label{fig:cmframe}
\end{figure}

We analyze the TCS process in both the $\gamma d$ and
$l^{+}l^{-}$ center-of-mass (c.m.) frames, as illustrated in
Fig.~\ref{fig:cmframe}. The dilepton c.m. frame is obtained from the
$ \gamma d$ c.m. frame by a Lorentz boost along the momentum of the
outgoing virtual photon. In the dilepton c.m. frame, $\theta$ denotes
the polar angle of the negatively charged lepton with momentum
$k_{1}$, while $\phi$ is the azimuthal angle between the lepton and
hadron scattering planes.

Denoting the magnitude of the transverse component of the momentum
transfer by
$\Delta_{T}= \left | \vec{\Delta}  _{T}  \right | $, its relation to the scattering
angle $\Theta$ in the $\gamma d$ c.m. frame is
\begin{equation}
  \sin \mathit{\Theta }=\frac{2\Delta_{T}\sqrt{s} }{r},
\end{equation}
where
\begin{equation}
r=\sqrt{(s-Q^{2}-m^{2})^{2}-4Q^{2}m^{2}}.
\end{equation}

In the dilepton c.m. frame, the lepton momenta can be written as
\begin{equation}
    k^{\mu}_{1}=\frac{Q}{2}\big(1,\beta\sin{\theta}\cos{\phi},\beta\sin{\theta} \sin{\phi},\beta\cos{\theta}\big),\ k^{\mu}_{2}=\frac{Q}{2}\big(1,-\beta\sin{\theta}\cos{\phi},-\beta\sin{\theta} \sin{\phi},-\beta\cos{\theta}\big),
\end{equation}
where $\beta$ is the lepton velocity in the dilepton c.m. frame,
\begin{equation}
\beta=\sqrt{1-\frac{4m_{l}^{2}}{Q^{2}}}.
\end{equation}

\begin{figure}[htbp]
\centering
\includegraphics[width=0.4\textwidth]{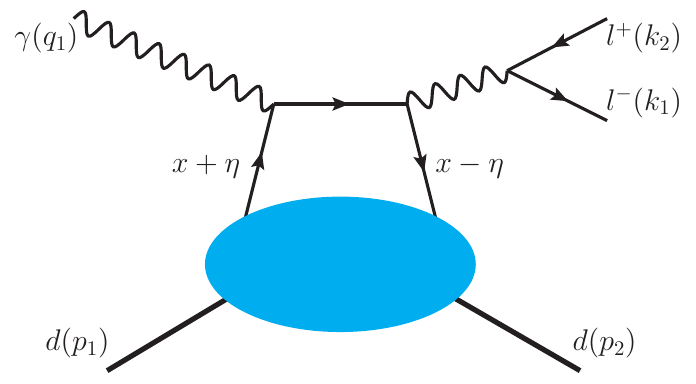}
\caption{A representative leading-order handbag diagram for TCS}
\label{fig:LOTCS}
\end{figure}

Two mechanisms contribute to the reaction
$\gamma d\to l^{+}l^{-}d$: the TCS and BH
processes. We first consider TCS,
\begin{equation}
\gamma+d\to\gamma^{\ast}+d\to l^{+}l^{-}+d,
\end{equation}
where the charge conjugation of the lepton pair is odd.
A representative leading-order handbag diagram for the TCS process is shown
Fig.~\ref{fig:LOTCS}. The incoming and outgoing partons carry
plus momentum fractions $x+\eta$ and $x-\eta$, respectively, with
respect to the average deuteron momentum $\bar{P}$. The skewness
parameter is defined as~\cite{Berger:2001xd}
\begin{equation}
\eta
=
-\frac{(q_{1}-q_{2})\cdot(q_{1}+q_{2})}
{(p_{1}+p_{2})\cdot(q_{1}+q_{2})}
\approx
\frac{Q^{2}}{2s-Q^{2}},
\end{equation}
where the approximate expression is obtained in the generalized
Bjorken limit after neglecting hadron-mass and momentum-transfer
corrections.

In the generalized
Bjorken limit, the TCS amplitude is expressed as a
convolution of perturbatively calculable hard scattering coefficient
functions with nonperturbative hadronic matrix elements parametrized
in terms of GPDs within collinear QCD factorization. For a spin-1 target, the twist-2 vector and
axial-vector quark GPDs are defined by~\cite{Berger:2001zb}
\begin{equation}
\begin{aligned}
&\frac{1}{2}\int\frac{dz^{-}}{2\pi}e^{ix\bar{P}^{+}z^{-}}\!\left\langle p_{2}\right| \bar{q} (-\frac{1 }{2 } z^{-})\gamma^{+}q(\frac{1 }{2 }  z^{-})\left | p_{1} \right \rangle \\=&-\left ( \epsilon_{2}^{\ast }  \cdot \epsilon_{1}   \right ) {H}_{1} +\frac{(\epsilon_{1}  \cdot q_{1})(\epsilon_{2}^{\ast }  \cdot P) +( \epsilon_{2}^{\ast }\cdot q_{1})( \epsilon_{1} \cdot P)}{P\cdot q_{1} }{H}_{2} -\frac{(\epsilon _{1}\cdot P )(\epsilon _{2}^{\ast } \cdot P)}{2m^{2} }{H} _{3} \\& + \frac{(\epsilon_{1}  \cdot q_{1})(\epsilon_{2}^{\ast }  \cdot P) -( \epsilon_{2}^{\ast }\cdot q_{1})( \epsilon_{1} \cdot P)}{P\cdot q_{1} }{H}_{4}  +\bigg [ 4m^{2} \frac{(\epsilon_{1}  \cdot q_{1})(\epsilon _{2}^{\ast }\cdot q_{1}  ) }{(P\cdot q_{1} )^{2} } +\frac{1}{3} (\epsilon _{2}^{\ast }\cdot \epsilon _{1}  )   \bigg ] {H}_{5},
\end{aligned}
\label{eqn:vgpds}
\end{equation}
\begin{equation}
\begin{aligned}
&\frac{1}{2}\int\frac{dz^{-}}{2\pi}e^{ix\bar{P}^{+}z^{-}}\!\left\langle p_{2}\right| \bar{q} (-\frac{1 }{2 } z^{-})\gamma^{+}\gamma _{5}q(\frac{1 }{2 }  z^{-})\left | p_{1} \right \rangle \\=&\ i\frac{\varepsilon _{q_{1} \epsilon _{2}^{\ast } \epsilon _{1}P }    }{P\cdot q_{1} } {\widetilde{H}  } _{1} -i\frac{\varepsilon _{q_{1} \Delta P\gamma } }{P\cdot q_{1} } \frac{\epsilon _{1}^{\gamma }(\epsilon _{2}^{\ast } \cdot P) +\epsilon _{2}^{\ast \gamma }(\epsilon _{1} \cdot P) }{m^{2} }{\widetilde{H} } _{2}-i\frac{\varepsilon _{q_{1} \Delta P\gamma } }{P\cdot q_{1} } \frac{\epsilon _{1}^{\gamma }(\epsilon _{2}^{\ast } \cdot P) -\epsilon _{2}^{\ast \gamma }(\epsilon _{1} \cdot P) }{m^{2} }{\widetilde{H} } _{3}\\ &-i \frac{\varepsilon _{q_{1} \Delta P\gamma } }{P\cdot q_{1} } \frac{\epsilon _{1}^{\gamma }(\epsilon _{2}^{\ast } \cdot q_{1} ) +\epsilon _{2}^{\ast \gamma }(\epsilon _{1} \cdot q_{1} ) }{P\cdot q_{1}  }{\widetilde{H} } _{4}.
\end{aligned}
\label{eqn:agpds}
\end{equation}
Here, $\epsilon_{1}$ and $\epsilon_{2}$ denote the
polarization vectors of the incoming and outgoing deuterons,
respectively. The flavor indices of the GPDs are suppressed for simplicity, and the gauge links required to ensure gauge invariance of the bilocal operators are understood. The GPDs are real-valued functions of the longitudinal momentum fraction $x$, the skewness parameter $\eta$, and the momentum transfer $t$, i.e., $H_i=H_i(x,\eta,t)$ and $\widetilde{H}_i=\widetilde{H}_i(x,\eta,t)$. Their dependence on the factorization scale $\mu_F$ is also suppressed throughout. In the forward limit, $H_1(x,0,0)$, $\widetilde{H}_1(x,0,0)$, and $H_5(x,0,0)$ reduce to the leading-twist unpolarized, helicity, and tensor-polarized PDFs, respectively.

\begin{figure}[htbp]
\centering
\includegraphics[width=0.4\textwidth]{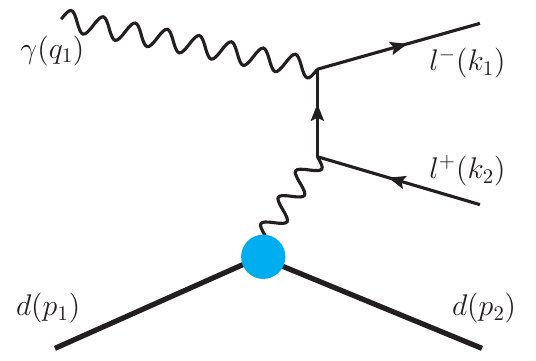}
\caption{One of the Feynman diagram for the BH progress}
\label{fig:BH}
\end{figure}

The second contribution to the coherent reaction
$\gamma d\to l^{+}l^{-}d$ is the BH process,
illustrated in Fig.~\ref{fig:BH}. In this case, the hadronic
part of the amplitude is determined by the elastic electromagnetic
FFs of the deuteron. The matrix elements of the vector and
axial-vector quark currents can be parametrized in terms of
the corresponding deuteron FFs as~\cite{Berger:2001zb}
\begin{equation}
\begin{aligned}
    \left \langle p_{2}  |\bar{q} (0)\gamma^{\mu}q(0)| p_{1}  \right \rangle &=-G_{1}(t)(\epsilon^{\ast}_{2}\cdot \epsilon_{1})P^{\mu}+G_{2}(t)\big[\epsilon_{1}^{\mu}(\epsilon^{\ast}_{2}\cdot P)+\epsilon^{\ast\mu}_{2}(\epsilon_{1}\cdot P)\big]-G_{3}(t)(\epsilon_{1}\cdot P)(\epsilon^{\ast}_{2}\cdot P)\frac{P^{\mu}}{2m^{2}},\\
   \left \langle p_{2}  |\bar{q} (0)\gamma^{\mu}\gamma_{5}q(0)| p_{1}  \right \rangle &=i\widetilde{G}_{1}(t)\varepsilon^{\mu\alpha\beta\gamma}\epsilon^{\ast}_{2\alpha}\epsilon_{1\beta}P_{\gamma}-i\widetilde{G}_{2}(t)\varepsilon^{\mu\alpha\beta\gamma}\Delta_{\alpha}P_{\beta}\frac{\epsilon_{1\gamma}(\epsilon_{2}^{\ast}\cdot P)+\epsilon_{2\gamma}^{\ast}(\epsilon_{1}\cdot P)}{m^2}.
   \label{eqn:EMFFs}
\end{aligned}
\end{equation}
The first Mellin moments of the deuteron GPDs are related to the
corresponding vector (electromagnetic) and axial-vector FFs through the sum
rules
\begin{equation}
    \begin{array}{l}
\int_{-1}^{1} d x H_{i}(x, \eta, t)=G_{i}(t) \quad(i=1,2,3), \\
\int_{-1}^{1} d x \widetilde{H}_{i}(x, \eta, t)=\widetilde{G}_{i}(t) \quad(i=1,2), \\
\int_{-1}^{1} d x H_{i}(x, \eta, t)=0 \quad(i=4,5), \\
\int_{-1}^{1} d x \widetilde{H}_{i}(x, \eta, t)=0 \quad(i=3,4).
\end{array}
\end{equation}
The first moments of
$H_{4}$, $H_{5}$, $\widetilde{H}_{3}$, and
$\widetilde{H}_{4}$ vanish because no corresponding Lorentz
structures occur in the decomposition of the local vector and
axial-vector current matrix elements of Eq.~\eqref{eqn:EMFFs}.
The electromagnetic FFs $G_{1,2,3}$ are related to the conventional
deuteron charge, magnetic, and quadrupole FFs through~\cite{Gutsche:2015xva}
\begin{equation}
G_{C}(t)=G_{1}(t)-\frac{2}{3} \delta G_{Q}(t), \quad G_{M}(t)=G_{2}(t), \quad G_{Q}(t)=G_{1}(t)-G_{2}(t)+(1-\delta) G_{3}(t), \quad \delta=\frac{t}{4 m^{2}}.
\end{equation}


\section{Hadronic tensor and Cross sections}

The unpolarized differential cross section for
$\gamma d\to l^+ l^-  d$ is given by~\cite{Heller:2020lnm}
\begin{equation}
    \frac{d\sigma }{d\tau dtd\cos {\theta }d\phi } =\frac{1}{(2\pi )^{4} } \frac{s}{64} \frac{\beta}{(s-m^{2})^{2} }  \overline{\sum_{\mathrm{spin}}}\left |\mathcal{ M}  \right |^{2},
\end{equation}
where $\tau=Q^{2}/s$, and the total amplitude $\mathcal{M}$ is  the  sum of the BH and TCS amplitudes,
\begin{equation}
\mathcal{M}=\mathcal{M}_{\mathrm{BH}}+\mathcal{M}_{\mathrm{TCS}}.
\end{equation}
To obtain compact analytical expressions for the cross sections, we
adopt the kinematic approximation of Ref.~\cite{Berger:2001xd}, in
which  $t$, $m_l^2$, and $m^2$ are neglected wherever they are
power suppressed relative to $Q^2$ and $s$. Within this
approximation,
\begin{equation}
    r\simeq s-Q^{2},
    \qquad
    \Delta_{T}^{2}
    \simeq
    -\frac{m^{2}Q^{4}+st(s-Q^{2})}{s^{2}}.
\end{equation}
The product of the two BH lepton-propagator denominators then reduces
to
\begin{equation}
    L_{\mathrm{BH}}=(k_{1}-q_{1})^{2}(q_{1}-k_{2})^{2}\approx \frac{Q^{4}\sin^{2}{\theta}}{4}.
\end{equation}
Since $L_{\mathrm{BH}}$ vanishes as $ \sin\theta\to 0$, this
approximation breaks down in the very forward and backward lepton
regions. In the numerical analysis, these regions should
therefore be excluded by imposing an appropriate angular cut.

We first consider the BH contribution to the cross section.
The result can be expressed in terms of the deuteron electromagnetic FFs defined in Eq.~\eqref{eqn:EMFFs}:
\begin{equation}
\begin{aligned}
\frac{d\sigma_{\mathrm{BH}}}{d\tau dtd\cos {\theta }d\phi }=&\frac{\alpha^{3}_{\mathrm{em}}}{12\pi s}\frac{1}{-t}\frac{1+\cos^{2}{\theta}}{\sin^{2}{\theta}}\Bigg\{4(1-\delta)(G_{2})^{2}-\frac{\Delta_{T}^{2}}{\tau^{2}t}\bigg[4\Big(4(\delta-1)\delta+3\Big)(G_{1})^{2}\\
&+8(2\delta-1)\delta(G_{2})^{2}+16(\delta-1)^{2}\delta^{2}(G_{3})^{2}-16(2\delta-1)\delta(G_{1}G_{2})\\
&-16\Big(1+\delta(2\delta-3)\Big)\delta(G_{1}G_{3})+32(\delta-1)\delta^{2}(G_{2}G_{3})\bigg]\Bigg\},
\end{aligned}
\label{eqn:BHc}
\end{equation}
where the BH contribution does not depend on the azimuthal angle $\phi$.

For the TCS contribution, the relevant hadronic tensor is defined through the off-forward matrix element of the time-ordered product of two electromagnetic currents,
\begin{equation}
   \mathcal{T}_{\mu\nu}= i\int \mathrm{d}^{4}x \  e^{-iq_{1}\cdot x} \left \langle d(p_{2}) \left | T\{J^{\mathrm{em}}_{\mu}(x)J^{\mathrm{em}}_{\nu}(0)\} \right | d(p_{1}) \right \rangle.
\end{equation}
At leading twist and LO  in $\alpha_s$, the quark handbag contribution of Fig.~\ref{fig:LOTCS} conserves the photon
helicity. The corresponding hadronic tensor can be decomposed into
vector and axial-vector parts as
\begin{equation}
 \mathcal{T}^{\mu \nu }_{0} =-g_{T}^{\mu \nu } \mathcal{V}-i\varepsilon _{T} ^{\mu \nu } \mathcal{A},
\label{eqn:nftensor}
\end{equation}
where the amplitudes $\mathcal{V}$ and $\mathcal{A}$ are expressed in terms
of the vector and axial-vector CFFs,
\begin{equation}
\begin{aligned}
\mathcal{V}=&-\left ( \epsilon_{2}^{\ast }  \cdot \epsilon_{1}   \right ) \mathcal{H}_{1} +\frac{(\epsilon_{1}  \cdot q_{1})(\epsilon_{2}^{\ast }  \cdot P) +( \epsilon_{2}^{\ast }\cdot q_{1})( \epsilon_{1} \cdot P)}{P\cdot q_{1} }\mathcal{H}_{2} -\frac{(\epsilon _{1}\cdot P )(\epsilon _{2}^{\ast } \cdot P)}{2m^{2} }\mathcal{H} _{3} \\
& + \frac{(\epsilon_{1}  \cdot q_{1})(\epsilon_{2}^{\ast }  \cdot P) -( \epsilon_{2}^{\ast }\cdot q_{1})( \epsilon_{1} \cdot P)}{P\cdot q_{1} }\mathcal{H}_{4}  +\bigg [ 4m^{2} \frac{(\epsilon_{1}  \cdot q_{1})(\epsilon _{2}^{\ast }\cdot q_{1}  ) }{(P\cdot q_{1} )^{2} } +\frac{1}{3} (\epsilon _{2}^{\ast }\cdot \epsilon _{1}  )   \bigg ] \mathcal{H}_{5},\\
\mathcal{A}=&\ i\frac{\varepsilon _{q_{1} \epsilon _{2}^{\ast } \epsilon _{1}P }    }{P\cdot q_{1} } \mathcal{\widetilde{H}  } _{1} -i\frac{\varepsilon _{q_{1} \Delta P\gamma } }{P\cdot q_{1} } \frac{\epsilon _{1}^{\gamma }(\epsilon _{2}^{\ast } \cdot P) +\epsilon _{2}^{\ast \gamma }(\epsilon _{1} \cdot P) }{m^{2} }\mathcal{\widetilde{H} } _{2}-i\frac{\varepsilon _{q_{1} \Delta P\gamma } }{P\cdot q_{1} } \frac{\epsilon _{1}^{\gamma }(\epsilon _{2}^{\ast } \cdot P) -\epsilon _{2}^{\ast \gamma }(\epsilon _{1} \cdot P) }{m^{2} }\mathcal{\widetilde{H} } _{3}\\ &-i \frac{\varepsilon _{q_{1} \Delta P\gamma } }{P\cdot q_{1} } \frac{\epsilon _{1}^{\gamma }(\epsilon _{2}^{\ast } \cdot q_{1} ) +\epsilon _{2}^{\ast \gamma }(\epsilon _{1} \cdot q_{1} ) }{P\cdot q_{1}  }\mathcal{\widetilde{H} } _{4}.
\label{eqn:cffs}
\end{aligned}
\end{equation}

These CFFs are given by convolutions of the quark GPDs with the corresponding LO coefficient functions,
\begin{equation}
\begin{aligned}
 \mathcal{H}_{i}(\eta, t) &=\sum_{q}\int dx\big[ C_{0}^{q}(x,\eta)- C_{0}^{q}(-x,\eta) \big] H^{q}_{i}(x, \eta, t) \quad(i=1,2,3,4,5),\\ \ \widetilde{\mathcal{H}}_{i}(\eta, t) &=-\sum_{q}\int dx\big[\widetilde{C}_{0}^{q}(x,\eta)+\widetilde{C}_{0}^{q}(-x,\eta) \big] \widetilde{H} ^{q}_{i}(x, \eta, t) \quad(i=1,2,3,4).
\end{aligned}
\label{eqn:LOCFFs}
\end{equation}
The tree-level quark coefficient functions $C^{q}_{0}$ and $\widetilde{C}^{q}_{0}$ are simply given by~\cite{Berger:2001xd}
\begin{equation}
    C_{0}^{q}(x,\eta)=-e_{q}^{2}\frac{1}{x+\eta+i\varepsilon},\quad
     \widetilde{C}_{0}^{q}(x,\eta)=e_{q}^{2}\frac{1}{x+\eta+i\varepsilon}.
     \label{eqn:lohc}
\end{equation}

Using the hadronic tensor in Eq.~\eqref{eqn:nftensor}, we obtain the
TCS contribution to the unpolarized differential cross section,
\begin{equation}
\begin{aligned}
\frac{d\sigma_{\mathrm{TCS}}}{d\tau dtd\cos {\theta }d\phi } =&\frac{\alpha^{3}_{\mathrm{em}}}{24\pi sQ^{2}} (1+\cos^{2}{\theta})\Bigg\{\frac{1}{2}\big[4(\delta-1)\delta+3\big]|\mathcal{H}_{1}|^{2}-2\delta(2\delta-1)\mathrm{Re}(\mathcal{H}_{1}\mathcal{H}_{2}^{\ast})-2\delta(\delta-1)(2\delta-1)\\
&\times\mathrm{Re}(\mathcal{H}_{1}\mathcal{H}_{3}^{\ast})-2(\delta-1)(2\delta-1)\eta\mathrm{Re}(\mathcal{H}_{1}\mathcal{H}_{4}^{\ast})+\frac{1}{3}\Big[\big(9-6\delta\big)\eta^{2}+2\delta(5-2\delta)\Big]\mathrm{Re}(\mathcal{H}_{1}\mathcal{H}_{5}^{\ast})\\
&+4(\delta-1)\delta^{2}\mathrm{Re}(\mathcal{H}_{2}\mathcal{H}_{3}^{\ast})+4\delta(\delta-1)\eta \mathrm{Re}(\mathcal{H}_{2}\mathcal{H}_{4}^{\ast})+\frac{2}{3}(\delta-2)(2\delta+3\eta^{2})\mathrm{Re}(\mathcal{H}_{2}\mathcal{H}_{5}^{\ast})\\
&-\big[\eta^{2}-2\delta^{2}+\delta(1-\eta^{2})\big]|\mathcal{H}_{2}|^{2}+2\delta^{2}(\delta-1)^{2}|\mathcal{H}_{3}|^{2}+4(\delta-1)^{2}\delta\eta\mathrm{Re}(\mathcal{H}_{3}\mathcal{H}_{4}^{\ast})
\\
&+\frac{2}{3}\Big[3(\delta-1)^{2}\eta^{2}+\delta\big(2(\delta-3)\delta+1\big)\Big]\mathrm{Re}(\mathcal{H}_{3}\mathcal{H}_{5}^{\ast})+\frac{2}{3}\eta\Big[3(\delta-1)\eta^{2}+2\big((\delta-3)\delta-1\big)\Big]\\
&\times\mathrm{Re}(\mathcal{H}_{4}\mathcal{H}_{5}^{\ast})+\Big[\big(\delta-1\big)\big(2\delta-1\big)\eta^{2}-\delta\Big]|\mathcal{H}_{4}|^{2}+\frac{1}{18}\big[4\delta(\delta-4)+12\delta\eta^{2}-9\eta^{2}(4-\eta^{2})+6\big]|\mathcal{H}_{5}|^{2}\\
&-\frac{\Delta_{T}^{2}}{m^{2}}(1+\eta)^{2}\Big[2(\delta-1)\mathrm{Re}(\mathcal{\widetilde{H}}_{1}\mathcal{\widetilde{H}}_{2}^{\ast})-\frac{1}{2}\mathrm{Re}(\mathcal{\widetilde{H}}_{1}\mathcal{\widetilde{H}}_{4}^{\ast})+2\delta\mathrm{Re}(\mathcal{\widetilde{H}}_{2}\mathcal{\widetilde{H}}_{4}^{\ast})-4\delta(\delta-1)|\mathcal{\widetilde{H}}_{2}|^{2}\\&-2(\delta-1)\eta\mathrm{Re}(\mathcal{\widetilde{H}}_{3}\mathcal{{\widetilde{H}}}_{4}^{\ast})-4\delta(\delta-1)|\mathcal{\widetilde{H}}_{3}|^{2}-\frac{1}{4}(1+\eta^{2})|\mathcal{\widetilde{H}}_{4}|^{2}\Big]-(\delta-1)(1-\eta^{2})|\mathcal{\widetilde{H}}_{1}|^{2}\Bigg\}.
\end{aligned}
\label{eqn:TCSc1}
\end{equation}
All nine twist-2 quark GPDs of the deuteron defined in
Eqs.~\eqref{eqn:vgpds} and~\eqref{eqn:agpds} enter the cross section
through the corresponding CFFs. Compared with the BH contribution,
the pure TCS term is suppressed parametrically by a factor of
$(-t)/Q^{2}$. Consequently,  the TCS
contribution is typically smaller than the BH background
in the small-$|t|$ region relevant to collinear QCD factorization.
Within the leading-twist, leading-order approximation adopted here,
the pure TCS contribution is also independent of the azimuthal angle
$\phi$, as is the BH contribution in the present kinematic
approximation. A nontrivial azimuthal dependence may arise once
additional photon-helicity structures are included, for example from
higher-order QCD contributions involving gluon transversity GPDs,
higher-twist effects, or polarization of the incoming photon~\cite{Goritschnig:2014eba,Boer:2015fwa}.

The interference between the BH and TCS
amplitudes contributes to the unpolarized differential cross section as
\begin{equation}
\begin{aligned}
\frac{d\sigma_{\mathrm{INT}}}{d\tau dtd\cos {\theta }d\phi } =&\frac{\alpha_{\mathrm{em}}^{3}}{3\pi }\frac{\Delta_{T}}{Q^{3}t}\Big(\frac{1+\cos^{2}{\theta}}{\sin{\theta}}\cos{\phi}\Big)\mathrm{Re}\Bigg\{\frac{1}{2}\mathcal{H}_{1}\bigg[\Big(4(\delta-1)\delta+3\Big)G_{1}-2\delta(2\delta-1)\Big(G_{2}+(\delta-1)G_{3}\Big)\bigg]\\&+\delta\mathcal{H}_{2}\Big[(1-2\delta)G_{1}+(2\delta-1)G_{2}+2\delta(\delta-1)G_{3}\Big]+(\delta-1)\delta\mathcal{H}_{3}\bigg[(1-2\delta)G_{1}\\
&+2\delta\Big(G_{2}+(\delta-1)G_{3}\Big)\bigg]+(\delta-1)\eta\mathcal{H}_{4}\bigg[(1-2\delta)G_{1}+2\delta\Big(G_{2}+(\delta-1)G_{3}\Big)\bigg]\\
&-\frac{1}{3}\mathcal{H}_{5}\bigg[\Big(3(\delta-\frac{3}{2})\eta^{2}+(2\delta-5)\delta\Big)G_{1}-\Big(3(\delta-1)\eta^{2}+2(\delta-2)\delta\Big)G_{2}\\
& -\Big(3(\delta-1)^{2}\eta^{2}+\delta\big(2(\delta-3)\delta+1\big)\Big)G_{3}\bigg]+\eta G_{2}\Big[(\delta-1)\mathcal{\widetilde{H}}_{1}-4(\delta-1)\delta\mathcal{\widetilde{H}}_{2}+\delta\mathcal{\widetilde{H}}_{4}\Big]\Bigg\},
\end{aligned}
\label{eqn:INTc1}
\end{equation}
which exhibits the characteristic $\cos\phi$ harmonic also encountered
in TCS on the proton~\cite{Berger:2001xd}.

The interference term vanishes upon integration over the full azimuthal
range. To retain sensitivity to this contribution, and hence to the
deuteron GPDs, we exploit the transformation
$(\theta,\phi)\to(\pi-\theta,\pi+\phi)$. Under this transformation,
the BH and pure TCS contributions in Eqs.~\eqref{eqn:BHc}
and~\eqref{eqn:TCSc1} are invariant, whereas the interference term in
Eq.~\eqref{eqn:INTc1} changes sign:
 \begin{equation}
d\sigma_{\mathrm{BH}}\to d\sigma_{\mathrm{BH}}, \quad
d\sigma_{\mathrm{TCS}}  \to d\sigma_{\mathrm{TCS}}, \quad
d\sigma_{\mathrm{INT}}  \to -d\sigma_{\mathrm{INT}},
\label{eq:reps}
\end{equation}
which is a consequence of different charge conjugation of the lepton pairs in the two contributing processes.

This symmetry motivates the definition of a forward-backward
asymmetry in which the cross section integrated over
$\pi/2\leq\phi\leq3\pi/2$ at  polar and azimuthal angles ($\theta, \phi$) is compared with
that in the complementary  region at ($\pi -\theta,\pi+\phi$):
\begin{equation}
    A_{\mathrm{FB}} (\theta) =\frac{\int_{\pi/2}^{3\pi/2} d\phi\frac{d\sigma (\theta,\phi)}{d\cos \theta ~ d \phi}- \int_{3\pi/2}^{2\pi} d \phi\frac{d\sigma (\pi -\theta,\phi)}{d\cos \theta ~ d \phi}- \int_{0}^{\pi/2} d \phi\frac{d\sigma (\pi -\theta,\phi)}{d\cos \theta ~ d \phi}}{\int_{\pi/2}^{3\pi/2} d\phi\frac{d\sigma (\theta,\phi)}{d\cos \theta ~ d \phi}+ \int_{3\pi/2}^{2\pi} d \phi\frac{d\sigma (\pi -\theta,\phi)}{d\cos \theta ~ d \phi}+ \int_{0}^{\pi/2} d \phi\frac{d\sigma (\pi -\theta,\phi)}{d\cos \theta ~ d \phi}}\,.
    \label{eqn:fba}
\end{equation}
By construction, the BH and pure TCS contributions cancel in the
numerator of $A_{\mathrm{FB}}$, leaving only the BH--TCS interference
term. The corresponding numerator is
\begin{equation}
\begin{aligned}
\frac{d\sigma_{\mathrm{FB}}}{d\tau dt d\cos{\theta}}&=-\frac{4\alpha_{\mathrm{em}}^{3}}{3\pi}\frac{\Delta_{T}}{Q^{3}t}\Big(\frac{1+\cos^{2}{\theta}}{\sin{\theta}}\Big)\mathrm{Re}\Bigg\{\frac{1}{2}\mathcal{H}_{1}\bigg[\Big(4(\delta-1)\delta+3\Big)G_{1}-2\delta(2\delta-1)\Big(G_{2}+(\delta-1)G_{3}\Big)\bigg]\\&+\delta\mathcal{H}_{2}\Big[(1-2\delta)G_{1}+(2\delta-1)G_{2}+2\delta(\delta-1)G_{3}\Big]+(\delta-1)\delta\mathcal{H}_{3}\bigg[(1-2\delta)G_{1}\\
&+2\delta\Big(G_{2}+(\delta-1)G_{3}\Big)\bigg]+(\delta-1)\eta\mathcal{H}_{4}\bigg[(1-2\delta)G_{1}+2\delta\Big(G_{2}+(\delta-1)G_{3}\Big)\bigg]\\
&-\frac{1}{3}\mathcal{H}_{5}\bigg[\Big(3(\delta-\frac{3}{2})\eta^{2}+(2\delta-5)\delta\Big)G_{1}-\Big(3(\delta-1)\eta^{2}+2(\delta-2)\delta\Big)G_{2}\\
& -\Big(3(\delta-1)^{2}\eta^{2}+\delta\big(2(\delta-3)\delta+1\big)\Big)G_{3}\bigg]+\eta G_{2}\Big[(\delta-1)\mathcal{\widetilde{H}}_{1}-4(\delta-1)\delta\mathcal{\widetilde{H}}_{2}+\delta\mathcal{\widetilde{H}}_{4}\Big]\Bigg\}.
\end{aligned}
\label{eqn:sigmaFB}
\end{equation}
The forward-backward asymmetry therefore provides direct sensitivity
to deuteron GPDs through the BH-TCS interference, which has been used in the  recent measurement of proton TCS~\cite{CLAS:2021lky}. To disentangle the nine GPD contributions, one would need a complete set of measurements with fixed helicity states of the initial and final deuteron. This is not likely to be experimentally feasible in the near future. One may, however, note that at small $|t|$ and skewness $\eta$, kinematical factors favor the contribution to the forward-backward asymmetry which is proportional to $G_1 \mathcal{H}_{1}$; thus, in a first approximation, the $H_1$ GPD is  the easiest outcome of an experimental study.

The results presented above are obtained at LO in
$\alpha_s$. At the moderate values of $Q^2$ accessible at Jefferson
Lab, NLO QCD corrections may be important for
a quantitative description of deuteron TCS and for the extraction of
the corresponding GPDs. At NLO, the quark coefficient functions
receive one-loop corrections \cite{Pire:2011st}, while helicity-conserving gluon GPDs
contribute to the TCS amplitude for the first time through the
corresponding gluonic coefficient functions. In the flavor-singlet
sector, the quark and gluon GPDs are also coupled through QCD
evolution. These quark and gluon contributions conserve photon
helicity, as does the LO quark handbag amplitude.

\begin{figure}[htbp]
\centering
\includegraphics[width=0.4\textwidth]{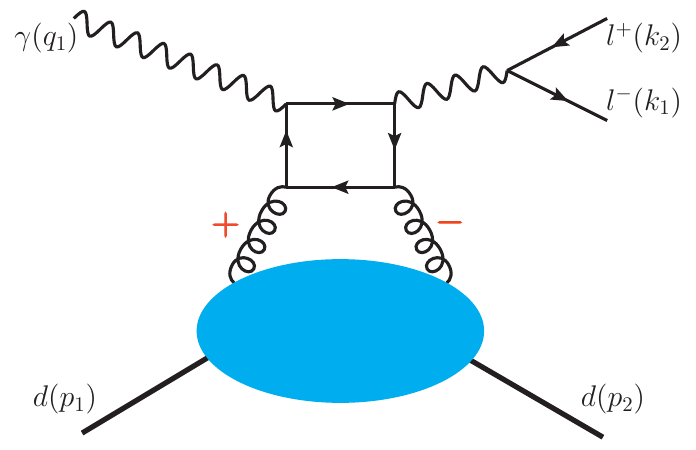}
\caption{A representative Feynman diagram contributing to the photon-helicity-flip TCS amplitude}
\label{fig:TCS-nlo}
\end{figure}

In addition to the photon-helicity-conserving terms, the gluon channel
also generates a photon-helicity-flip amplitude involving the gluon
transversity GPDs~\cite{Diehl:2001pm,Belitsky:2000jk,Hoodbhoy:1998vm}.
These distributions describe a two-unit helicity transfer in the gluon
GPDs and first contribute to the TCS amplitude at order $\alpha_s$.
In the partonic subprocess, the two-unit change in gluon helicity is
accompanied by a corresponding flip of the photon helicity, as
illustrated in Fig.~\ref{fig:TCS-nlo}. The interference of this
helicity-flip amplitude with the BH and helicity-conserving TCS amplitudes
generates characteristic azimuthal harmonics in the dilepton angular
distribution, providing a possible means of isolating the gluon
transversity contribution experimentally~\cite{Diehl:1997bu}. The NLO
hard-scattering coefficient functions and corrections to the cross sections are collected in Appendix~\ref{app-a}.

\section{Numerical Estimate OF THE CROSS SECTIONS}

In this section, we present LO phenomenological predictions for the
reaction
$\gamma d \to e^{+}e^{-}d$
using the deuteron electromagnetic FFs and GPDs as
nonperturbative inputs. Motivated by the experimental extraction of GPD information through the forward-backward asymmetry~\cite{CLAS:2021lky}, which involves integration over the azimuthal angle, we also integrate the BH and TCS cross sections over this angle in our numerical analysis.
We first specify the parametrizations adopted
for the deuteron FFs and GPDs.

For the deuteron electromagnetic FFs, we employ the analytic
soft-wall AdS/QCD parametrization of Ref.~\cite{Gutsche:2015xva}.
In this framework, the Fock component associated with the deuteron is
assigned twist $\tau=6$, and the three invariant electromagnetic FFs $G_{1,2,3}$ are expressed in terms of a universal twist-6
hadronic FF~\cite{Gutsche:2015xva},
\begin{equation}
    F(t)=\frac{\Gamma(6)\Gamma(a+1)}{\Gamma(a+6)}=\frac{5!}{(a+1)...(a+5)},
    \label{eqn:t6}
\end{equation}
where $a=-t/4\kappa^2$, and $\kappa=190\ \mathrm{MeV}$ is the scale parameter  that controls the $t$ dependence of the
FFs and is fixed from experimental measurements.

The deuteron electromagnetic FFs are given by
\begin{equation}
   G_{1}(t)=F(t),\quad G_{i}(t)=c_{i}F(t)\quad(i=2,3),
\end{equation}
The coefficients $c_{2}$ and $c_{3}$ are fixed by the normalization
of the magnetic and quadrupole FFs at zero momentum
transfer,
\begin{equation}
    c_{2}=G_{M}(0)=1.714,
    \qquad
    c_{3}=G_{M}(0)+G_{Q}(0)-1=26.544,
\end{equation}
where $G_{Q}(0)=25.83$. This parametrization provides a compact
description of the measured deuteron charge, magnetic, and quadrupole
FFs and exhibits the asymptotic behavior
$F(t)\sim 1/(|t|)^{5}$ expected from the quark-counting rules.

\begin{figure}[htb]
\centering
\includegraphics[width=0.6\textwidth]{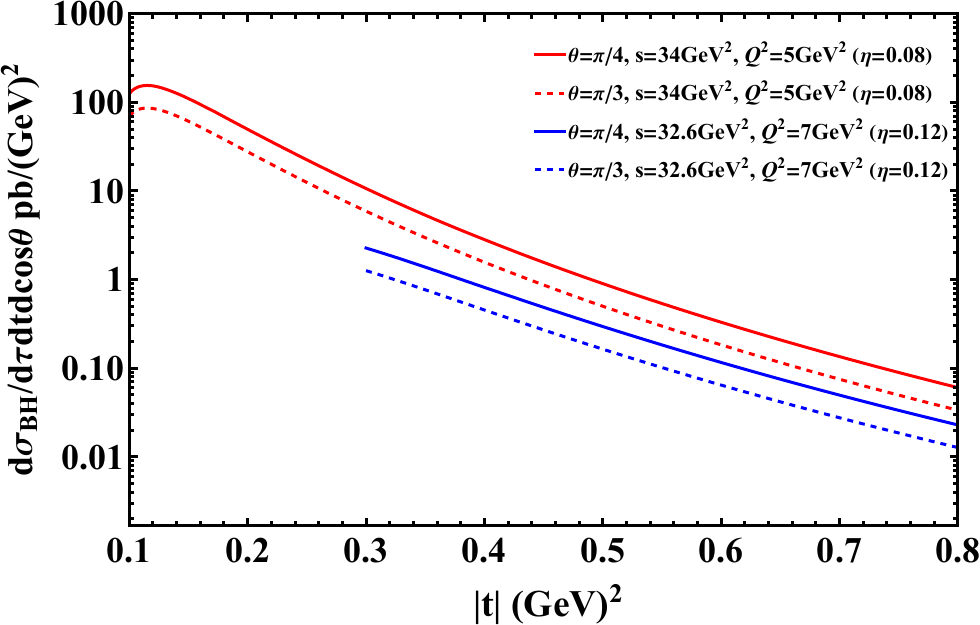}
\caption{The differential cross section of the BH process}
\label{fig:estBH}
\end{figure}

For the deuteron GPDs defined in
Eqs.~\eqref{eqn:vgpds} and~\eqref{eqn:agpds}, we employ  a convolution
model in which the hard interaction takes place on one of the
constituent nucleons, while the second nucleon acts as a spectator~\cite{Cano:2002ph,Cosyn:2018rdm}.
The deuteron GPDs are thus obtained by convoluting the corresponding
nucleon GPDs with the deuteron light-cone wave function. For the
nucleon GPDs, we use the Goloskokov--Kroll (GK) parametrization
~\cite{Goloskokov:2006hr,Goloskokov:2008ib,Goloskokov:2009ia},
whereas the deuteron wave function is based on the
Argonne~$v_{18}$ potential of Ref.~\cite{Wiringa:1994wb}.
This construction incorporates the dominant proton--neutron
structure of the deuteron and provides the nonperturbative input for
our numerical evaluation of the TCS amplitude.

\begin{figure}[htb]
\centering
\includegraphics[width=0.6\textwidth]{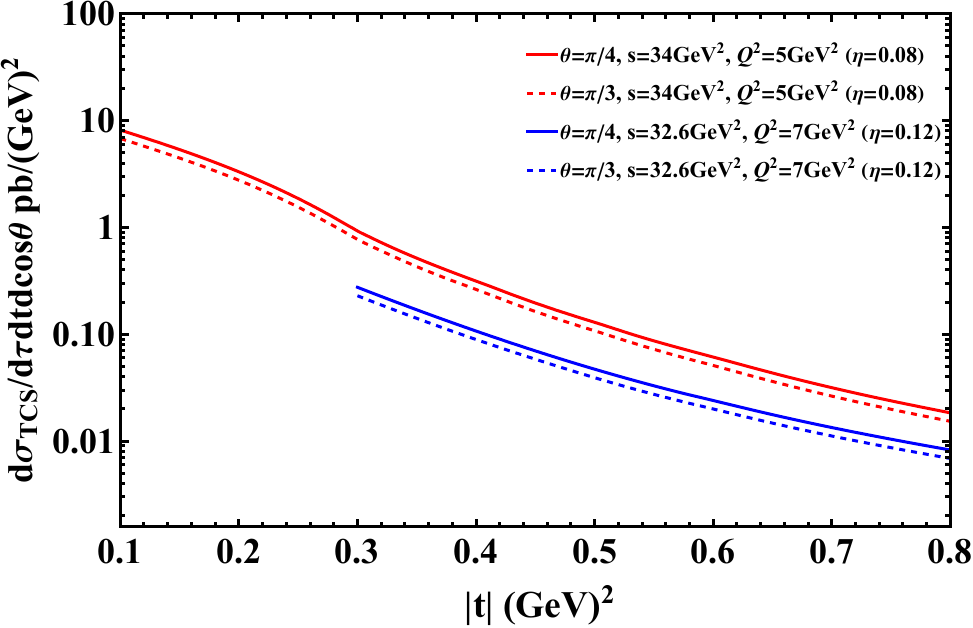}
\caption{The differential cross section of the TCS process}
\label{fig:estTCS}
\end{figure}

For TCS, we restrict the dilepton invariant mass to the range
$2~\mathrm{GeV}<Q<3~\mathrm{GeV}$. This choice provides a compromise
between maintaining a sufficiently large hard scale for the applicability
of collinear factorization and avoiding prominent hadronic resonance
regions. It lies within the so-called ``resonance-free window'' commonly
adopted in TCS phenomenology and is therefore favorable for the extraction
of deuteron GPDs from future measurements.

\begin{figure}[htb]
\centering
\includegraphics[width=0.6\textwidth]{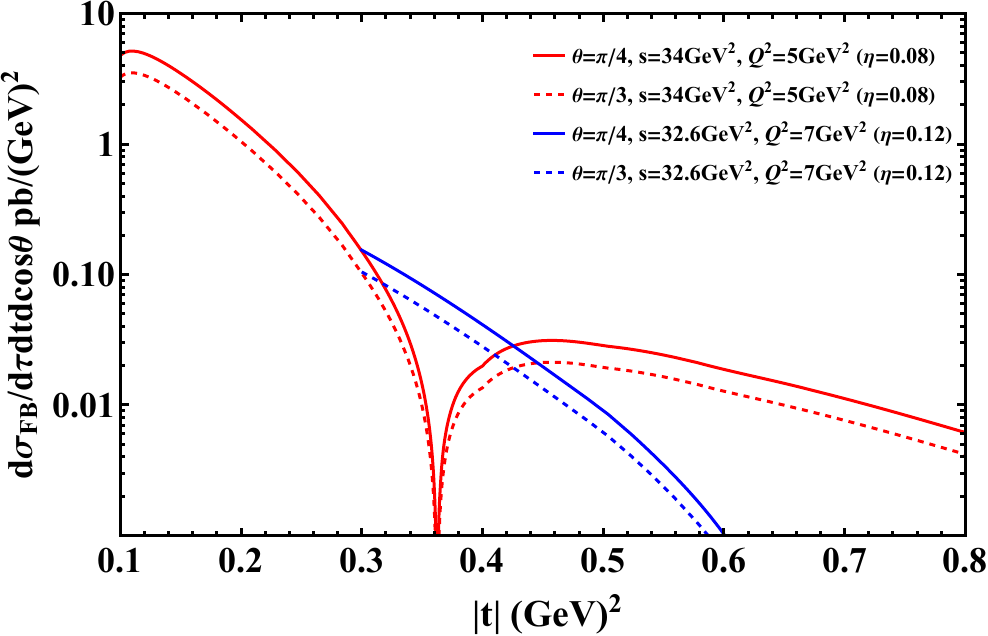}
\caption{The estimate of the interference
contribution through the absolute value of} the forward-backward asymmetry
\label{fig:estsigmaFB}
\end{figure}

In the existing proton TCS measurements at Jefferson Lab, the incident
photon energy is typically in the range
$E_{\gamma}\simeq 7$--$9~\mathrm{GeV}$~\cite{CLAS:2021lky}.
For a deuteron target at rest, this corresponds to a squared
c.m. energy of approximately
$30~\mathrm{GeV}^{2}<s<37~\mathrm{GeV}^{2}$. In the kinematic region
considered here, the corresponding skewness parameter is approximately $\eta\simeq 0.1$.

In Fig.~\ref{fig:estBH}, we show the $t$ dependence of the BH differential
cross section evaluated using the deuteron electromagnetic FFs.
The different colors correspond to different values of the skewness
parameter $\eta$. The choice $\eta=0.08$ corresponds to the kinematic
point $s=34~\mathrm{GeV}^{2}$ and $Q^{2}=5~\mathrm{GeV}^{2}$, whereas
$\eta=0.12$ corresponds to $s=32.6~\mathrm{GeV}^{2}$ and
$Q^{2}=7~\mathrm{GeV}^{2}$. Different values of $\eta$ also lead to
different kinematic lower limits on $|t|$,
\begin{equation}
|t|_{\mathrm{min}}=\frac{s-m^{2}}{2s}\Bigg[s+Q^{2}-m^{2}-\sqrt{\lambda(s,Q^{2},m^{2})}\Bigg]-Q^{2},
\end{equation}
with
$\lambda(s,Q^{2},m^{2})=(s-Q^{2}-m^{2})^{2}-4m^{2}Q^{2}$.
The solid and dashed curves in this figure
denote $\theta=\pi/4$ and $\theta=\pi/3$, respectively. Owing to the
form-factor behavior specified in Eq.~\eqref{eqn:t6}, the BH contribution
decreases rapidly with increasing $|t|$.

Fig.~\ref{fig:estTCS} shows the $t$ dependence of the pure TCS
contribution evaluated using the deuteron GPDs obtained within the
convolution model. We adopt the same line conventions and kinematic
settings as in Fig.~\ref{fig:estBH}. As expected, the pure TCS
contribution is approximately one order of magnitude smaller than the
BH contribution over the kinematic region considered.

In Fig.~\ref{fig:estsigmaFB}, we show the BH-TCS interference
contribution through the absolute value of the
forward--backward observable defined in
Eq.~\eqref{eqn:sigmaFB}, since this observable can change sign.
Comparison with Fig.~\ref{fig:estTCS} shows that the interference
contribution is comparable in magnitude to the pure TCS contribution.
Its magnitude decreases rapidly with increasing $|t|$.The forward-backward observable changes sign at
$|t|\approx 0.35~\mathrm{GeV}^2$ for $\eta=0.08$ and
$|t|\approx 0.64~\mathrm{GeV}^2$ for $\eta=0.12$,
giving rise to the corresponding dips in the logarithmic plot.
After these sign changes, its magnitude remains small over
the $t$ range considered.
Given the presently limited constraints on deuteron GPDs,
however, the numerical results for both the pure TCS and
interference contributions should be regarded primarily as
order-of-magnitude estimates rather than precision predictions.

\section{Summary}

In this work, we have investigated coherent TCS on a deuteron target as a probe of  the deuteron GPDs. Owing to its spin-1 nature, the deuteron exhibits a considerably richer GPD structure than the nucleon.
We have presented the formalism for the TCS and BH amplitudes for a spin-1 target and derived the corresponding differential cross sections. Numerical estimates at LO were obtained in the kinematic region relevant to future measurements at Jefferson Lab. The deuteron quark GPDs were evaluated within a convolution model, while the BH contribution was calculated using a phenomenological parametrization of the deuteron electromagnetic FFs. We studied the relative magnitudes and kinematic dependences of the TCS, BH, and interference contributions. These results provide quantitative guidance for identifying kinematic regions in which the TCS signal and its interference with the BH process can be experimentally accessed.

Given the moderate energy scale accessible in future measurements at Jefferson Lab, higher-order QCD corrections~\cite{Moutarde:2013qs,Boussarie:2023xun} may play an important role in the corresponding cross sections. We have therefore also presented the NLO formalism for deuteron TCS, with particular emphasis on contributions involving the gluon transversity GPDs. These distributions encode helicity-flip gluonic structure and give rise to characteristic angular dependences in the c.m. frame of the produced lepton pair. Such angular structures can, in principle, be exploited to isolate the gluon transversity contribution from the dominant BH and LO TCS cross sections. A quantitative numerical analysis of the NLO corrections is left for future work, pending the development of a sufficiently reliable model for the gluon GPDs of the deuteron.

Our results demonstrate that coherent deuteron TCS can serve as a complementary probe of both quark and gluon GPDs in a spin-1 system. Future measurements at Jefferson Lab would provide valuable information on the multidimensional partonic structure of the deuteron and may help clarify the interplay between conventional nucleonic degrees of freedom, tensor polarization, and possible non-nucleonic components of the deuteron.

\section*{Acknowledgments}
We are indebted to Prof.~Wim Cosyn  for providing the grid of the deuteron GPDs.
Qin-Tao Song was supported by the Natural Science Foundation of Henan Province under Grant  No. 252300423011.
Bo-Qiang Ma was supported by the National Natural Science Foundation of China (Grants No.~12675108 and No.~12335006).

\appendix
\section{NLO corrections for TCS with gluon transversity GPDs}
\label{app-a}

Including the NLO QCD corrections, the photon-helicity-conserving CFFs of Eq.~\eqref{eqn:cffs} are expressed in terms of quark and gluon GPDs,
\begin{equation}
\begin{aligned}
&\mathcal{H}_{i}(\eta,t)=\int_{-1}^{1}  d x\Big( \sum_{q}T^{q}(x,\eta)H^{q}_{i}(x,\eta,t)+T^{g}(x,\eta)H^{g}_{i}(x,\eta,t)\Big)\quad(i=1,2,3,4,5), \\
&\widetilde{\mathcal{H}}_{i}(\eta,t)=- \int_{-1}^{1}  d x\left(\sum_{q}\widetilde{T}^{q}(x,\eta)\widetilde{H}^{q}_{i}(x,\eta,t)+\widetilde{T}^{g}(x,\eta)\widetilde{H}^{g}_{i}(x,\eta,t)\right)\quad(i=1,2,3,4).
\label{eqn:CFFs}
\end{aligned}
\end{equation}
Here, the coefficient functions $T$ and $\widetilde T$ contain the LO
and NLO contributions. Up to NLO accuracy, they are given by
\begin{equation}
\begin{aligned}
T^{q}(x)&=\Big[C^{q}_{0}(x)+C^{q}_{1}(x)\Big]-(x\to-x),\quad T^{g}(x)=C^{g}_{1}(x)+(x\to-x),\\ \widetilde{T}^{q}(x)&=\Big[\widetilde{C}^{q}_{0}(x)+\widetilde{C}^{q}_{1}(x)\Big]+(x\to-x),\quad \widetilde{T}^{g}(x)=\widetilde{C}^{g}_{1}(x)-(x\to-x).
\end{aligned}
\label{eqn:cofunctions1}
\end{equation}

To complete the description of the photon-helicity-conserving amplitude
given in the main text, we collect here the NLO coefficient functions
entering Eq.~\eqref{eqn:cofunctions1}. The functions
$C_{1}^{q,g}$ and $\widetilde{C}_{1}^{q,g}$ describe the one-loop
corrections in the quark and gluon channels, respectively, and read
~\cite{Pire:2011st,Moutarde:2013qs}
\begin{equation}
\begin{aligned}
    C_{1}^{q}(x,\eta)=&\frac{e_{q}^{2}\alpha_{s}C_{F}}{4\pi}\frac{1}{x+\eta+i\varepsilon}\Bigg\{9-\pi^{2}+3i\pi-3\frac{x+\eta}{x-\eta}\log\Big(\frac{x+\eta}{2\eta}+i\varepsilon \Big)-\bigg[\log\Big(\frac{x+\eta}{2\eta}+i\varepsilon\Big)-i\pi\bigg]^{2}\Bigg\},\\
    C_{1}^{g}(x,\eta)=&\frac{\sum_{q}e_{q}^{2}\alpha_{s}T_{F}}{4\pi}\frac{1}{(x+\eta+i\varepsilon)(x-\eta-i\varepsilon)}\Bigg\{2\frac{x+3\eta}{x-\eta}\log\Big(\frac{x+\eta}{2\eta}+i\varepsilon\Big)\\&-\frac{x+\eta}{x-\eta}\bigg[\pi^{2}+\bigg(\log\Big(\frac{x+\eta}{2\eta}+i\varepsilon\Big)-i\pi\bigg)^{2}\bigg]\Bigg\},\\
    \widetilde{C}_{1}^{q}(x,\eta)=&-\frac{e_{q}^{2}\alpha_{s}C_{F}}{4\pi}\frac{1}{x+\eta+i\varepsilon}\Bigg\{9-\pi^{2}+3i\pi-\frac{x+\eta}{x-\eta}\log\Big(\frac{x+\eta}{2\eta}+i\varepsilon \Big)-\bigg[\log\Big(\frac{x+\eta}{2\eta}+i\varepsilon\Big)-i\pi\bigg]^{2}\Bigg\},\\
    \widetilde{C}_{1}^{g}(x,\eta)=&-\frac{\sum_{q}e_{q}^{2}\alpha_{s}T_{F}}{4\pi}\frac{1}{(x+\eta+i\varepsilon)(x-\eta-i\varepsilon)}\Bigg\{-2\frac{3x+\eta}{x-\eta}\log\Big(\frac{x+\eta}{2\eta}+i\varepsilon\Big)\\&+\frac{x+\eta}{x-\eta}\bigg[\pi^{2}+\bigg(\log\Big(\frac{x+\eta}{2\eta}+i\varepsilon\Big)-i\pi\bigg)^{2}\bigg]\Bigg\},
\end{aligned}
\end{equation}
with
\begin{equation}
C_F=\frac{N_c^2-1}{2N_c},
\qquad
T_F=\frac{1}{2},
\end{equation}
and the factorization scale has been chosen as $\mu_F=Q$.

At NLO, the hadronic tensor in Eq.~\eqref{eqn:nftensor} receives an
additional tensor contribution associated with the photon-helicity-flip
amplitude,
\begin{equation}
     \mathcal{T}^{\mu\nu}= \mathcal{T}^{\mu\nu}_{0}+ \mathcal{T}^{\mu\nu}_{1},
     \label{eqn:htto}
\end{equation}
where $\mathcal{T}^{\mu\nu}_{0}$ denotes the photon-helicity-conserving
contribution, whereas $\mathcal{T}^{\mu\nu}_{1}$ describes the
photon-helicity-flip amplitude.
The latter can be expressed in terms of the twist-2 gluon transversity
GPDs of the deuteron, which are defined through the off-forward matrix
element of the gluon tensor operator~\cite{Cosyn:2018rdm},
\begin{equation}
\begin{aligned}
    F_{T}^{\mu\nu}=&-\frac{2}{P^{+}}\int \frac{d z^{-}}{2\pi}e^{ix\bar{P}^{+}z^{-}}\left \langle p_{2},\lambda_{2}\right| \mathrm{Tr} \mathbf{\hat{S} } [G^{+\mu}(-\frac{1}{2}z^{-})G^{+\nu}(\frac{1}{2}z^{-})]\left|  p_{1},\lambda_{1}\right \rangle \\
    =&\mathbf{\hat{S} }\Bigg\{2(\Delta+\eta P)^{\mu} \frac{(\epsilon_{2}^{\ast}\cdot q_{1})\epsilon_{1}^{\nu}-\epsilon_{2}^{\ast\nu}(\epsilon_{1}\cdot q_{1})}{(P\cdot q_{1})}H^{gT}_{1}+2(\Delta+\eta P)^{\mu} \bigg[\frac{2P^{\nu}(\epsilon_{1}\cdot q_{1})(\epsilon_{2}^{\ast}\cdot q_{1})}{(P\cdot q_{1})^{2}}-\frac{(\epsilon_{1}\cdot q_{1})\epsilon_{2}^{\ast\nu}+\epsilon_{1}^{\nu}(\epsilon_{2}^{\ast}\cdot q_{1})}{(P\cdot q_{1})}\bigg]H^{gT}_{2}\\
    &+\frac{(\Delta+\eta P)^{\mu}}{m}\bigg[\frac{(\epsilon_{2}^{\ast}\cdot q_{1})\Delta^{\nu}-\epsilon_{2}^{\ast\nu}(\Delta\cdot  q_{1})}{m(P\cdot q_{1})}(\epsilon_{1}\cdot P)-\frac{(\epsilon_{1}\cdot q_{1})\Delta^{\nu}-\epsilon_{1}^{\nu}(\Delta\cdot q_{1})}{m(P\cdot q_{1})}(\epsilon_{2}^{\ast}\cdot P)\bigg]H^{gT}_{3}\\
    &+\frac{(\Delta+\eta P)^{\mu}}{m}\bigg[\frac{(\epsilon_{2}^{\ast}\cdot q_{1})\Delta^{\nu}-\epsilon_{2}^{\ast\nu}(\Delta\cdot  q_{1})}{m(P\cdot q_{1})}(\epsilon_{1}\cdot P)+\frac{(\epsilon_{1}\cdot q_{1})\Delta^{\nu}-\epsilon_{1}^{\nu}(\Delta\cdot q_{1})}{m(P\cdot q_{1})}(\epsilon_{2}^{\ast}\cdot P)\bigg]H^{gT}_{4}\\
    &-\bigg[\frac{(\epsilon_{2}^{\ast}\cdot q_{1})P^{\mu}-(P\cdot q_{1})\epsilon_{2}^{\ast\mu}}{(P\cdot q_{1})}\bigg]\bigg[\frac{(\epsilon_{1}\cdot q_{1})P^{\nu}-(P\cdot q_{1})\epsilon_{1}^{\nu}}{(P\cdot q_{1})}\bigg]H_{5}^{gT}\\&+\bigg[\frac{(\epsilon_{2}^{\ast}\cdot q_{1})\Delta^{\mu}-(\Delta\cdot q_{1})\epsilon_{2}^{\ast\mu}}{(P\cdot q_{1})}\bigg]\bigg[\frac{(\epsilon_{1}\cdot q_{1})\Delta^{\nu}-(\Delta\cdot q_{1})\epsilon_{1}^{\nu}}{(P\cdot q_{1})}\bigg]H_{6}^{gT} +\frac{(\Delta+\eta P)^{\mu}}{m}\frac{(\Delta+\eta P)^{\nu}}{m}\frac{(\epsilon_{2}^{\ast}\cdot P)(\epsilon_{1}\cdot P)}{4m^{2}}H_{7}^{gT}\\&+\frac{(\Delta+\eta P)^{\mu}}{m}\bigg[\frac{(\epsilon_{2}^{\ast}\cdot q_{1})P^{\nu}-\epsilon_{2}^{\ast\nu}(P\cdot q_{1})}{2m(P\cdot q_{1})}(\epsilon_{1}\cdot P)+\frac{(\epsilon_{1}\cdot q_{1})P^{\nu}-\epsilon_{1}^{\nu}(P\cdot q_{1})}{2m(P\cdot q_{1})}(\epsilon_{2}^{\ast}\cdot P)\bigg]H^{gT}_{8}
    \\&+\frac{(\Delta+\eta P)^{\mu}}{m}\bigg[\frac{(\epsilon_{2}^{\ast}\cdot q_{1})P^{\nu}-\epsilon_{2}^{\ast\nu}(P\cdot q_{1})}{2m(P\cdot q_{1})}(\epsilon_{1}\cdot P)-\frac{(\epsilon_{1}\cdot q_{1})P^{\nu}-\epsilon_{1}^{\nu}(P\cdot q_{1})}{2m(P\cdot q_{1})}(\epsilon_{2}^{\ast}\cdot P)\bigg]H^{gT}_{9}\Bigg\},
\end{aligned}
\label{eqn:gTGPDs}
\end{equation}
here $\mathbf{\hat{S}}$ denotes symmetrization in the indices
$\mu$ and $\nu$ together with subtraction of the trace. Explicitly,
\begin{equation}
    \mathbf{\hat{S}}[A^{\mu}B^{\nu}]
    \equiv
    \tau_{T}^{\mu\nu;\rho\sigma}A_{\rho}B_{\sigma},
\end{equation}
with the transverse symmetric-traceless projector~\cite{Belitsky:2001ns}
\begin{equation}
    \tau_{T}^{\mu\nu;\rho\sigma}
    =
    \frac{1}{2}
    \left(
    g_{T}^{\mu\rho}g_{T}^{\nu\sigma}
    +g_{T}^{\mu\sigma}g_{T}^{\nu\rho}
    -g_{T}^{\mu\nu}g_{T}^{\rho\sigma}
    \right).
\end{equation}
In the forward limit, only the gluon transversity GPD
$H_{5}^{gT}(x,0,0)$ remains nonzero. It reduces to the gluon
transversity PDF of a spin-1 hadron
~\cite{Jaffe:1989xy,Nzar:1992ax,Kumano:2020gfk,Kumano:2019igu,
Winter:2017bfs,Maxwell:2018gci}.

The photon-helicity-flip hadronic tensor
$\mathcal{T}_{1}^{\mu\nu}$ has the same Lorentz decomposition as
Eq.~\eqref{eqn:gTGPDs}, with the gluon transversity GPDs
$H_{i}^{gT}$ replaced by the corresponding CFFs
$\mathcal{H}_{i}^{gT}$. These CFFs are given by
~\cite{Hoodbhoy:1998vm}
\begin{equation}
    \mathcal{H}_{i}^{gT}(\eta,t)
    =
    \frac{\alpha_{s}}{2\pi}
    \sum_{q}e_{q}^{2}
    \int_{-1}^{1}dx\,
    \frac{H_{i}^{gT}(x,\eta,t)}
    {(x+\eta+i\varepsilon)(x-\eta-i\varepsilon)},
    \qquad (i=1,\ldots,9) .
    \label{eqn:gTCFFs}
\end{equation}
Together with the photon-helicity-conserving contribution, this
completes the NLO expression for the hadronic tensor in Eq.~\eqref{eqn:htto}.

At NLO, the TCS amplitude can be decomposed into the
photon-helicity-conserving and photon-helicity-flip contributions,
\begin{equation}
     \mathcal{M}_ {\mathrm{TCS}}= \mathcal{M}^{0}_ {\mathrm{TCS}}+\mathcal{M}^{1}_{\mathrm{TCS}},
     \label{eqn:tcsnlo}
\end{equation}
in direct correspondence with the decomposition of the hadronic tensor
in Eq.~\eqref{eqn:htto}. Accordingly, the pure TCS cross section can be
written as
 \begin{equation}
d\sigma_{\mathrm{TCS}}=d\sigma^0_{\mathrm{TCS}}+d\sigma^1_{\mathrm{TCS}}+d\sigma^2_{\mathrm{TCS}}.
\label{eq:tcsnlo1}
\end{equation}
where $d\sigma^0_{\mathrm{TCS}}$ denotes the contribution from the
photon-helicity-conserving amplitude. It is obtained from
Eq.~\eqref{eqn:TCSc1} by replacing the LO CFFs with the corresponding
NLO expressions given in Eq.~\eqref{eqn:CFFs}. The second term,
$d\sigma^1_{\mathrm{TCS}}$, arises from the interference between the
photon-helicity-conserving and photon-helicity-flip TCS amplitudes.
The last term, $d\sigma^2_{\mathrm{TCS}}$, originates from the squared
photon-helicity-flip amplitude,
$|\mathcal{M}^{1}_{\mathrm{TCS}}|^{2}$. Since
$\mathcal{M}^{1}_{\mathrm{TCS}}$ starts at order $\alpha_s$, this term
is of order $\alpha_s^2$ and is therefore beyond NLO accuracy. It is
neglected in the following.

Using the explicit expression for
$\mathcal{M}^{1}_{\mathrm{TCS}}$, the interference contribution
$d\sigma^1_{\mathrm{TCS}}$ is given by
\begin{equation}
	\begin{aligned}
		\frac{d\sigma^{1}_{\mathrm{TCS}}}{d\tau dtd\cos {\theta }d\phi } =&\frac{\alpha^{3}_{\mathrm{em}}}{48\pi s Q^{2}}\frac{\Delta_{T}^{2}}{m^{2}}(\cos{2\phi}\sin^{2}{\theta})(1+\eta)^{2}\mathrm{Re}\Bigg\{\mathcal{H}_{1}^{gT}\Big[(2\delta-1)(\mathcal{H}_{1}-\mathcal{H}_{2})^{\ast}-2(\delta-1)\delta\mathcal{H}_{3}^{\ast}\\&-(2\delta-1)\eta\mathcal{H}_{4}^{\ast}-\frac{1}{3}(3\eta^{2}+2\delta-4)\mathcal{H}_{5}^{\ast}+\eta\tilde{\mathcal{H}}^{\ast}_{1}-4(\delta-1)\eta\widetilde{\mathcal{H}}^{\ast}_{2}
		+4\delta\widetilde{\mathcal{H}}^{\ast}_{3}+2\eta\widetilde{\mathcal{H}}^{\ast}_{4}\Big]\\&+\mathcal{H}_{2}^{gT}\Big[(2\delta-3)\eta(\mathcal{H}_{1}-\mathcal{H}_{2})^{\ast}-2(\delta-1)^{2}\eta\mathcal{H}_{3}^{\ast}-\big(2(\delta-1)\eta^{2}-1\big)\mathcal{H}_{4}^{\ast}\\
		&-\frac{1}{3}\eta\big(3\eta^{2}+2(\delta-3)\big)\mathcal{H}_{5}^{\ast}+\widetilde{\mathcal{H}}^{\ast}_{1}-4\delta\widetilde{\mathcal{H}}^{\ast}_{2}+4(\delta-1)\eta\widetilde{\mathcal{H}}^{\ast}_{3}+(1+\eta^{2})\widetilde{\mathcal{H}}^{\ast}_{4}\Big]\\&-2\mathcal{H}_{3}^{gT}\Big[\delta\mathcal{H}^{\ast}_{4}+ \eta\mathcal{H}^{\ast}_{5}-4(\delta-1)\delta\eta\widetilde{\mathcal{H}}^{\ast}_{3}-(\delta-1)\eta^{2}\widetilde{\mathcal{H}}^{\ast}_{4}\Big]\\&+2\mathcal{H}_{4}^{gT}\Big[\delta(2\delta-1)(\mathcal{H}_{1}-\mathcal{H}_{2})^{\ast} -2(\delta-1)\delta^{2}\mathcal{H}_{3}^{\ast}-2(\delta-1)\delta\eta\mathcal{H}_{4}^{\ast}\\&-\frac{1}{3}\big(3(\delta-1)\eta^{2}+2(\delta-2)\delta\big)\mathcal{H}_{5}^{\ast}+(\delta-1)\eta\widetilde{\mathcal{H}}^{\ast}_{1}-4(\delta-1)\delta\eta\tilde{\mathcal{H}}^{\ast}_{2}+\delta\eta\widetilde{\mathcal{H}}^{\ast}_{4}\Big] \\&+\mathcal{H}_{5}^{gT}\Big[\frac{1}{4}(3-2\delta)\mathcal{H}_{1}^{\ast}+\frac{1}{2}(\delta-1)\mathcal{H}_{2}^{\ast}+\frac{1}{2}(\delta-1)^{2}\mathcal{H}_{3}^{\ast}+\frac{1}{2}\eta(\delta-1)\mathcal{H}_{4}^{\ast}\\&+\frac{1}{12}\big(3\eta^{2}+2(\delta-3)\big)\mathcal{H}_{5}^{\ast}-2(\delta-1)\widetilde{\mathcal{H}}^{\ast}_{3}-\frac{1}{2}\eta\widetilde{\mathcal{H}}^{\ast}_{4}\Big]-\frac{1}{4}\mathcal{H}_{6}^{gT}\Big[(2\delta+1)\mathcal{H}_{1}^{\ast}\\&-2\delta(\mathcal{H}_{2}+\delta\mathcal{H}_{3}+\eta\mathcal{H}_{4})^{\ast}-\frac{1}{3}\big(3\eta^{2}+2(\delta-1)\big)\mathcal{H}_{5}^{\ast}+2\eta\widetilde{\mathcal{H}}_{1}^{\ast}-8\eta\delta\widetilde{\mathcal{H}}_{2}^{\ast}+2\eta\widetilde{\mathcal{H}}_{4}^{\ast} \Big]\\&+\mathcal{H}_{7}^{gT}\Big[(1-2\delta)(\delta-1)\delta\mathcal{H}_{1}^{\ast}+2(\delta-1)\delta^{2}\mathcal{H}_{2}^{\ast}+2(\delta-1)^{2}\delta^{2}\mathcal{H}_{3}^{\ast}+2(\delta-1)^{2}\delta\eta\mathcal{H}_{4}^{\ast}\\&+\frac{1}{3}\big(\delta+2(\delta-3)\delta^{2}+3(\delta-1)^{2}\eta^{2}\big)\mathcal{H}_{5}^{\ast}\Big]+\mathcal{H}_{8}^{gT}\Big[(1-\delta)\eta\mathcal{H}_{2}^{\ast}+\eta\mathcal{H}_{5}^{\ast}-(\delta-1)\widetilde{\mathcal{H}}_{1}^{\ast}\\&+4(\delta-1)\delta\widetilde{\mathcal{H}}_{2}^{\ast}-\delta\widetilde{\mathcal{H}}_{4}^{\ast}\Big]+\mathcal{H}_{9}^{gT}\Big[(1-\delta)(2\delta-1)\mathcal{H}_{1}^{\ast}+2(\delta-1)\delta\mathcal{H}_{2}^{\ast}+2(\delta-1)^{2}\delta\mathcal{H}_{3}^{\ast}\\&+(2\delta-1)(\delta-1)\eta\mathcal{H}_{4}^{\ast}+\frac{1}{3}\big(1+2\delta(\delta-3)+3(\delta-1)\eta^{2}\big)\mathcal{H}_{5}^{\ast}-4(\delta-1)\delta\widetilde{\mathcal{H}}_{3}^{\ast}-(\delta-1)\eta\widetilde{\mathcal{H}}_{4}^{\ast}\Big]\Bigg\},
        \label{eqn:tsc-tr}
	\end{aligned}
\end{equation}
where the explicit CFFs are given in Eqs.~\eqref{eqn:CFFs} and  \eqref{eqn:gTCFFs}. This contribution
generates a characteristic $\cos(2\phi)$ harmonic in the pure TCS
cross section, in contrast to the azimuthally independent
photon-helicity-conserving contribution
$d\sigma^0_{\mathrm{TCS}}$.

The BH-TCS interference contribution can similarly be decomposed as
\begin{equation}
d\sigma_{\mathrm{INT}}=d\sigma^0_{\mathrm{INT}}+d\sigma^1_{\mathrm{INT}},
\label{eq:tcsnlo}
\end{equation}
where $d\sigma^0_{\mathrm{INT}}$ denotes the interference between the
BH amplitude and the photon-helicity-conserving TCS amplitude. It is
obtained from Eq.~\eqref{eqn:INTc1} by replacing the LO CFFs with the
NLO expressions in Eq.~\eqref{eqn:CFFs}. The second term,
$d\sigma^1_{\mathrm{INT}}$, arises from the interference between the BH
amplitude and the photon-helicity-flip TCS amplitude,
\begin{equation}
    \begin{aligned}
\frac{d\sigma^{1}_{\mathrm{INT}}}{d\tau dtd\cos {\theta }d\phi } =&  \frac{\alpha_{\mathrm{em}}^{3}}{12\pi} \frac{\Delta^{3}_{T}}{m^{2}Q^{3}t}(\cos{3\phi}\sin{\theta})(1+\eta)^{2} \mathrm{Re}\Bigg\{\mathcal{H}_{1}^{gT}\Big[(2\delta-1)(G_{1}-G_{2})-2(\delta-1)\delta G_{3}\Big]\\&+\mathcal{H}_{2}^{gT}\Big[(2\delta-3)\eta G_{1}-2(\delta-1)\eta G_{2}-2(\delta-1)^{2}\eta G_{3}\Big]+\mathcal{H}_{4}^{gT}\Big[2(2\delta-1)\delta (G_{1}-G_{2})\\&-4(\delta-1)\delta^{2}G_{3}\Big]+\frac{1}{4}\mathcal{H}_{5}^{gT}\Big[(3-2\delta)G_{1}+2(\delta-1)G_{2}+2(\delta-1)^{2}G_{3}\Big]\\&-\frac{1}{4}\mathcal{H}_{6}^{gT}\Big[(2\delta+1)G_{1}-2\delta G_{2}-2\delta^{2}G_{3}\Big]+\mathcal{H}_{7}^{gT}(\delta-1)\delta\Big[(1-2\delta)G_{1}+2\delta G_{2}+2(\delta-1)\delta G_{3}\Big]\\&+\mathcal{H}_{9}^{gT}(\delta-1)\Big[(1-2\delta)G_{1}+2\delta G_{2}+2(\delta-1)\delta G_{3}\Big]\Bigg\},
   \label{eqn:tsc-int}
    \end{aligned}
\end{equation}
which involves the CFFs associated with the gluon transversity GPDs
and generates a characteristic $\cos(3\phi)$ harmonic.

We also consider the NLO corrections to the forward--backward
asymmetry defined in Eq.~\eqref{eqn:fba}. At this order, the
denominator retains the same form as in the LO calculation, with
the LO CFFs replaced by their NLO counterparts. The gluon
transversity contributions do not enter the denominator at
this order because their azimuthal harmonics vanish upon
integration over the corresponding angular regions. In contrast, the interference term  $d\sigma^{1}_{\mathrm{INT}}$ gives an additional contribution
to the numerator. Accordingly, we write
\begin{equation}
d\sigma_{\mathrm{FB}}=d\sigma^0_{\mathrm{FB}}+d\sigma^1_{\mathrm{FB}},
\label{eq:fbnlo}
\end{equation}
where $d\sigma^0_{\mathrm{FB}}$ is obtained from Eq.~\eqref{eqn:sigmaFB}
 by replacing the LO CFFs with the corresponding
NLO ones, and the additional contribution $d\sigma^1_{\mathrm{FB}}$ is given by
\begin{equation}
    \begin{aligned}
\frac{d\sigma^{1}_{\mathrm{FB}}}{d\tau dtd\cos {\theta }} =&  \frac{\alpha_{\mathrm{em}}^{3}}{9\pi } \frac{\Delta^{3}_{T}}{m^{2}Q^{3}t}\sin{\theta}(1+\eta)^{2} \mathrm{Re}\Bigg\{\mathcal{H}_{1}^{gT}\Big[(2\delta-1)(G_{1}-G_{2})-2(\delta-1)\delta G_{3}\Big]\\&+\mathcal{H}_{2}^{gT}\Big[(2\delta-3)\eta G_{1}-2(\delta-1)\eta G_{2}-2(\delta-1)^{2}\eta G_{3}\Big]+\mathcal{H}_{4}^{gT}\Big[2(2\delta-1)\delta (G_{1}-G_{2})\\&-4(\delta-1)\delta^{2}G_{3}\Big]+\frac{1}{4}\mathcal{H}_{5}^{gT}\Big[(3-2\delta)G_{1}+2(\delta-1)G_{2}+2(\delta-1)^{2}G_{3}\Big]\\&-\frac{1}{4}\mathcal{H}_{6}^{gT}\Big[(2\delta+1)G_{1}-2\delta G_{2}-2\delta^{2}G_{3}\Big]+\mathcal{H}_{7}^{gT}(\delta-1)\delta\Big[(1-2\delta)G_{1}+2\delta G_{2}+2(\delta-1)\delta G_{3}\Big]\\&+\mathcal{H}_{9}^{gT}(\delta-1)\Big[(1-2\delta)G_{1}+2\delta G_{2}+2(\delta-1)\delta G_{3}\Big]\Bigg\}.
   \label{eqn:tsc-fba1}
    \end{aligned}
\end{equation}
This observable therefore provides
a complementary means of probing the gluon transversity GPDs of the deuteron, for which a quantitative prediction awaits the development of a dedicated gluon GPD model.

Although the gluon-transversity contributions in
Eqs.~\eqref{eqn:tsc-tr} and~\eqref{eqn:tsc-int} first appear at order
$\alpha_s$, their characteristic azimuthal harmonics provide a means
of separating them from the dominant photon-helicity-conserving
contributions through an angular analysis in the c.m. frame
of the produced lepton pair. These observables therefore provide a
potentially useful probe of the gluon transversity GPDs of the
deuteron.




\bibliography{bibliography}

\begin{thebibliography}{86}%
\makeatletter
\providecommand \@ifxundefined [1]{%
 \@ifx{#1\undefined}
}%
\providecommand \@ifnum [1]{%
 \ifnum #1\expandafter \@firstoftwo
 \else \expandafter \@secondoftwo
 \fi
}%
\providecommand \@ifx [1]{%
 \ifx #1\expandafter \@firstoftwo
 \else \expandafter \@secondoftwo
 \fi
}%
\providecommand \natexlab [1]{#1}%
\providecommand \enquote  [1]{``#1''}%
\providecommand \bibnamefont  [1]{#1}%
\providecommand \bibfnamefont [1]{#1}%
\providecommand \citenamefont [1]{#1}%
\providecommand \href@noop [0]{\@secondoftwo}%
\providecommand \href [0]{\begingroup \@sanitize@url \@href}%
\providecommand \@href[1]{\@@startlink{#1}\@@href}%
\providecommand \@@href[1]{\endgroup#1\@@endlink}%
\providecommand \@sanitize@url [0]{\catcode `\\12\catcode `\$12\catcode
  `\&12\catcode `\#12\catcode `\^12\catcode `\_12\catcode `\%12\relax}%
\providecommand \@@startlink[1]{}%
\providecommand \@@endlink[0]{}%
\providecommand \url  [0]{\begingroup\@sanitize@url \@url }%
\providecommand \@url [1]{\endgroup\@href {#1}{\urlprefix }}%
\providecommand \urlprefix  [0]{URL }%
\providecommand \Eprint [0]{\href }%
\providecommand \doibase [0]{https://doi.org/}%
\providecommand \selectlanguage [0]{\@gobble}%
\providecommand \bibinfo  [0]{\@secondoftwo}%
\providecommand \bibfield  [0]{\@secondoftwo}%
\providecommand \translation [1]{[#1]}%
\providecommand \BibitemOpen [0]{}%
\providecommand \bibitemStop [0]{}%
\providecommand \bibitemNoStop [0]{.\EOS\space}%
\providecommand \EOS [0]{\spacefactor3000\relax}%
\providecommand \BibitemShut  [1]{\csname bibitem#1\endcsname}%
\let\auto@bib@innerbib\@empty
\bibitem [{\citenamefont {Hoodbhoy}\ \emph {et~al.}(1989)\citenamefont
  {Hoodbhoy}, \citenamefont {Jaffe},\ and\ \citenamefont
  {Manohar}}]{Hoodbhoy:1988am}%
  \BibitemOpen
  \bibfield  {author} {\bibinfo {author} {\bibfnamefont {P.}~\bibnamefont
  {Hoodbhoy}}, \bibinfo {author} {\bibfnamefont {R.~L.}\ \bibnamefont
  {Jaffe}},\ and\ \bibinfo {author} {\bibfnamefont {A.}~\bibnamefont
  {Manohar}},\ }\bibfield  {title} {\bibinfo {title} {{Novel Effects in Deep
  Inelastic Scattering from Spin 1 Hadrons}},\ }\href
  {https://doi.org/10.1016/0550-3213(89)90572-5} {\bibfield  {journal}
  {\bibinfo  {journal} {Nucl. Phys. B}\ }\textbf {\bibinfo {volume} {312}},\
  \bibinfo {pages} {571} (\bibinfo {year} {1989})}\BibitemShut {NoStop}%
\bibitem [{\citenamefont {Close}\ and\ \citenamefont
  {Kumano}(1990)}]{Close:1990zw}%
  \BibitemOpen
  \bibfield  {author} {\bibinfo {author} {\bibfnamefont {F.~E.}\ \bibnamefont
  {Close}}\ and\ \bibinfo {author} {\bibfnamefont {S.}~\bibnamefont {Kumano}},\
  }\bibfield  {title} {\bibinfo {title} {{A sum rule for the spin dependent
  structure function b-1(x) for spin one hadrons}},\ }\href
  {https://doi.org/10.1103/PhysRevD.42.2377} {\bibfield  {journal} {\bibinfo
  {journal} {Phys. Rev. D}\ }\textbf {\bibinfo {volume} {42}},\ \bibinfo
  {pages} {2377} (\bibinfo {year} {1990})}\BibitemShut {NoStop}%
\bibitem [{\citenamefont {Kumano}(2010)}]{Kumano:2010vz}%
  \BibitemOpen
  \bibfield  {author} {\bibinfo {author} {\bibfnamefont {S.}~\bibnamefont
  {Kumano}},\ }\bibfield  {title} {\bibinfo {title} {{Tensor-polarized quark
  and antiquark distribution functions in a spin-one hadron}},\ }\href
  {https://doi.org/10.1103/PhysRevD.82.017501} {\bibfield  {journal} {\bibinfo
  {journal} {Phys. Rev. D}\ }\textbf {\bibinfo {volume} {82}},\ \bibinfo
  {pages} {017501} (\bibinfo {year} {2010})},\ \Eprint
  {https://arxiv.org/abs/1005.4524} {arXiv:1005.4524 [hep-ph]} \BibitemShut
  {NoStop}%
\bibitem [{\citenamefont {Kumano}\ and\ \citenamefont
  {Song}(2016)}]{Kumano:2016ude}%
  \BibitemOpen
  \bibfield  {author} {\bibinfo {author} {\bibfnamefont {S.}~\bibnamefont
  {Kumano}}\ and\ \bibinfo {author} {\bibfnamefont {Q.-T.}\ \bibnamefont
  {Song}},\ }\bibfield  {title} {\bibinfo {title} {{Theoretical estimate on
  tensor-polarization asymmetry in proton-deuteron Drell-Yan process}},\ }\href
  {https://doi.org/10.1103/PhysRevD.94.054022} {\bibfield  {journal} {\bibinfo
  {journal} {Phys. Rev. D}\ }\textbf {\bibinfo {volume} {94}},\ \bibinfo
  {pages} {054022} (\bibinfo {year} {2016})},\ \Eprint
  {https://arxiv.org/abs/1606.03149} {arXiv:1606.03149 [hep-ph]} \BibitemShut
  {NoStop}%
\bibitem [{\citenamefont {Airapetian}\ \emph {et~al.}(2005)\citenamefont
  {Airapetian} \emph {et~al.}}]{HERMES:2005pon}%
  \BibitemOpen
  \bibfield  {author} {\bibinfo {author} {\bibfnamefont {A.}~\bibnamefont
  {Airapetian}} \emph {et~al.} (\bibinfo {collaboration} {HERMES}),\ }\bibfield
   {title} {\bibinfo {title} {{First measurement of the tensor structure
  function b(1) of the deuteron}},\ }\href
  {https://doi.org/10.1103/PhysRevLett.95.242001} {\bibfield  {journal}
  {\bibinfo  {journal} {Phys. Rev. Lett.}\ }\textbf {\bibinfo {volume} {95}},\
  \bibinfo {pages} {242001} (\bibinfo {year} {2005})},\ \Eprint
  {https://arxiv.org/abs/hep-ex/0506018} {arXiv:hep-ex/0506018} \BibitemShut
  {NoStop}%
\bibitem [{\citenamefont {Khan}\ and\ \citenamefont
  {Hoodbhoy}(1991)}]{Khan:1991qk}%
  \BibitemOpen
  \bibfield  {author} {\bibinfo {author} {\bibfnamefont {H.}~\bibnamefont
  {Khan}}\ and\ \bibinfo {author} {\bibfnamefont {P.}~\bibnamefont
  {Hoodbhoy}},\ }\bibfield  {title} {\bibinfo {title} {{Convenient
  parametrization for deep inelastic structure functions of the deuteron}},\
  }\href {https://doi.org/10.1103/PhysRevC.44.1219} {\bibfield  {journal}
  {\bibinfo  {journal} {Phys. Rev. C}\ }\textbf {\bibinfo {volume} {44}},\
  \bibinfo {pages} {1219} (\bibinfo {year} {1991})}\BibitemShut {NoStop}%
\bibitem [{\citenamefont {Cosyn}\ \emph {et~al.}(2017)\citenamefont {Cosyn},
  \citenamefont {Dong}, \citenamefont {Kumano},\ and\ \citenamefont
  {Sargsian}}]{Cosyn:2017fbo}%
  \BibitemOpen
  \bibfield  {author} {\bibinfo {author} {\bibfnamefont {W.}~\bibnamefont
  {Cosyn}}, \bibinfo {author} {\bibfnamefont {Y.-B.}\ \bibnamefont {Dong}},
  \bibinfo {author} {\bibfnamefont {S.}~\bibnamefont {Kumano}},\ and\ \bibinfo
  {author} {\bibfnamefont {M.}~\bibnamefont {Sargsian}},\ }\bibfield  {title}
  {\bibinfo {title} {{Tensor-polarized structure function $b_1$ in standard
  convolution description of deuteron}},\ }\href
  {https://doi.org/10.1103/PhysRevD.95.074036} {\bibfield  {journal} {\bibinfo
  {journal} {Phys. Rev. D}\ }\textbf {\bibinfo {volume} {95}},\ \bibinfo
  {pages} {074036} (\bibinfo {year} {2017})},\ \Eprint
  {https://arxiv.org/abs/1702.05337} {arXiv:1702.05337 [hep-ph]} \BibitemShut
  {NoStop}%
\bibitem [{\citenamefont {Miller}(2014)}]{Miller:2013hla}%
  \BibitemOpen
  \bibfield  {author} {\bibinfo {author} {\bibfnamefont {G.~A.}\ \bibnamefont
  {Miller}},\ }\bibfield  {title} {\bibinfo {title} {{Pionic and Hidden-Color,
  Six-Quark Contributions to the Deuteron b1 Structure Function}},\ }\href
  {https://doi.org/10.1103/PhysRevC.89.045203} {\bibfield  {journal} {\bibinfo
  {journal} {Phys. Rev. C}\ }\textbf {\bibinfo {volume} {89}},\ \bibinfo
  {pages} {045203} (\bibinfo {year} {2014})},\ \Eprint
  {https://arxiv.org/abs/1311.4561} {arXiv:1311.4561 [nucl-th]} \BibitemShut
  {NoStop}%
\bibitem [{\citenamefont {Farooq}(2026)}]{Farooq:2026kky}%
  \BibitemOpen
  \bibfield  {author} {\bibinfo {author} {\bibfnamefont {M.}~\bibnamefont
  {Farooq}},\ }\bibfield  {title} {\bibinfo {title} {{The $b_1$ (E12-13-011)
  Polarized Target Experiment}},\ }\href {https://doi.org/10.22323/1.517.0154}
  {\bibfield  {journal} {\bibinfo  {journal} {PoS}\ }\textbf {\bibinfo {volume}
  {SPIN2025}},\ \bibinfo {pages} {154} (\bibinfo {year} {2026})}\BibitemShut
  {NoStop}%
\bibitem [{\citenamefont {Keller}(2022)}]{Keller:2022abm}%
  \BibitemOpen
  \bibfield  {author} {\bibinfo {author} {\bibfnamefont {D.}~\bibnamefont
  {Keller}} (\bibinfo {collaboration} {SpinQuest}),\ }\bibfield  {title}
  {\bibinfo {title} {{The Transverse Structure of the Deuteron with
  Drell-Yan}},\ }\href@noop {} {\  (\bibinfo {year} {2022})},\ \Eprint
  {https://arxiv.org/abs/2205.01249} {arXiv:2205.01249 [nucl-ex]} \BibitemShut
  {NoStop}%
\bibitem [{\citenamefont {Clement}\ and\ \citenamefont
  {Keller}(2023)}]{Clement:2023eun}%
  \BibitemOpen
  \bibfield  {author} {\bibinfo {author} {\bibfnamefont {J.}~\bibnamefont
  {Clement}}\ and\ \bibinfo {author} {\bibfnamefont {D.}~\bibnamefont
  {Keller}},\ }\bibfield  {title} {\bibinfo {title} {{Manipulation of spin-1
  solid-state targets}},\ }\href {https://doi.org/10.1016/j.nima.2023.168177}
  {\bibfield  {journal} {\bibinfo  {journal} {Nucl. Instrum. Meth. A}\ }\textbf
  {\bibinfo {volume} {1050}},\ \bibinfo {pages} {168177} (\bibinfo {year}
  {2023})}\BibitemShut {NoStop}%
\bibitem [{\citenamefont {Poudel}\ \emph {et~al.}(2025)\citenamefont {Poudel},
  \citenamefont {Bacchetta}, \citenamefont {Chen},\ and\ \citenamefont
  {Santiesteban}}]{Poudel:2025nof}%
  \BibitemOpen
  \bibfield  {author} {\bibinfo {author} {\bibfnamefont {J.}~\bibnamefont
  {Poudel}}, \bibinfo {author} {\bibfnamefont {A.}~\bibnamefont {Bacchetta}},
  \bibinfo {author} {\bibfnamefont {J.-P.}\ \bibnamefont {Chen}},\ and\
  \bibinfo {author} {\bibfnamefont {N.}~\bibnamefont {Santiesteban}},\
  }\bibfield  {title} {\bibinfo {title} {{Experimental study of tensor
  structure function of deuteron}},\ }\href
  {https://doi.org/10.1140/epja/s10050-025-01558-w} {\bibfield  {journal}
  {\bibinfo  {journal} {Eur. Phys. J. A}\ }\textbf {\bibinfo {volume} {61}},\
  \bibinfo {pages} {81} (\bibinfo {year} {2025})},\ \Eprint
  {https://arxiv.org/abs/2506.04506} {arXiv:2506.04506 [nucl-ex]} \BibitemShut
  {NoStop}%
\bibitem [{\citenamefont {Arbuzov}\ \emph {et~al.}(2021)\citenamefont {Arbuzov}
  \emph {et~al.}}]{Arbuzov:2020cqg}%
  \BibitemOpen
  \bibfield  {author} {\bibinfo {author} {\bibfnamefont {A.}~\bibnamefont
  {Arbuzov}} \emph {et~al.},\ }\bibfield  {title} {\bibinfo {title} {{On the
  physics potential to study the gluon content of proton and deuteron at NICA
  SPD}},\ }\href {https://doi.org/10.1016/j.ppnp.2021.103858} {\bibfield
  {journal} {\bibinfo  {journal} {Prog. Part. Nucl. Phys.}\ }\textbf {\bibinfo
  {volume} {119}},\ \bibinfo {pages} {103858} (\bibinfo {year} {2021})},\
  \Eprint {https://arxiv.org/abs/2011.15005} {arXiv:2011.15005 [hep-ex]}
  \BibitemShut {NoStop}%
\bibitem [{\citenamefont {Koile}\ \emph {et~al.}(2012)\citenamefont {Koile},
  \citenamefont {Macaluso},\ and\ \citenamefont {Schvellinger}}]{Koile:2011aa}%
  \BibitemOpen
  \bibfield  {author} {\bibinfo {author} {\bibfnamefont {E.}~\bibnamefont
  {Koile}}, \bibinfo {author} {\bibfnamefont {S.}~\bibnamefont {Macaluso}},\
  and\ \bibinfo {author} {\bibfnamefont {M.}~\bibnamefont {Schvellinger}},\
  }\bibfield  {title} {\bibinfo {title} {{Deep Inelastic Scattering from
  Holographic Spin-One Hadrons}},\ }\href
  {https://doi.org/10.1007/JHEP02(2012)103} {\bibfield  {journal} {\bibinfo
  {journal} {JHEP}\ }\textbf {\bibinfo {volume} {02}},\ \bibinfo {pages}
  {103}},\ \Eprint {https://arxiv.org/abs/1112.1459} {arXiv:1112.1459 [hep-th]}
  \BibitemShut {NoStop}%
\bibitem [{\citenamefont {Detmold}(2006)}]{Detmold:2005iz}%
  \BibitemOpen
  \bibfield  {author} {\bibinfo {author} {\bibfnamefont {W.}~\bibnamefont
  {Detmold}},\ }\bibfield  {title} {\bibinfo {title} {{Target mass effects in
  deep-inelastic scattering on the deuteron}},\ }\href
  {https://doi.org/10.1016/j.physletb.2005.10.091} {\bibfield  {journal}
  {\bibinfo  {journal} {Phys. Lett. B}\ }\textbf {\bibinfo {volume} {632}},\
  \bibinfo {pages} {261} (\bibinfo {year} {2006})},\ \Eprint
  {https://arxiv.org/abs/hep-ph/0509011} {arXiv:hep-ph/0509011} \BibitemShut
  {NoStop}%
\bibitem [{\citenamefont {Kumano}\ and\ \citenamefont
  {Song}(2021{\natexlab{a}})}]{Kumano:2021fem}%
  \BibitemOpen
  \bibfield  {author} {\bibinfo {author} {\bibfnamefont {S.}~\bibnamefont
  {Kumano}}\ and\ \bibinfo {author} {\bibfnamefont {Q.-T.}\ \bibnamefont
  {Song}},\ }\bibfield  {title} {\bibinfo {title} {{Twist-2 relation and sum
  rule for tensor-polarized parton distribution functions of spin-1 hadrons}},\
  }\href {https://doi.org/10.1007/JHEP09(2021)141} {\bibfield  {journal}
  {\bibinfo  {journal} {JHEP}\ }\textbf {\bibinfo {volume} {09}},\ \bibinfo
  {pages} {141}},\ \Eprint {https://arxiv.org/abs/2106.15849} {arXiv:2106.15849
  [hep-ph]} \BibitemShut {NoStop}%
\bibitem [{\citenamefont {Kumano}\ and\ \citenamefont
  {Song}(2022)}]{Kumano:2021xau}%
  \BibitemOpen
  \bibfield  {author} {\bibinfo {author} {\bibfnamefont {S.}~\bibnamefont
  {Kumano}}\ and\ \bibinfo {author} {\bibfnamefont {Q.-T.}\ \bibnamefont
  {Song}},\ }\bibfield  {title} {\bibinfo {title} {{Equation-of-motion and
  Lorentz-invariance relations for tensor-polarized parton distribution
  functions of spin-1 hadrons}},\ }\href
  {https://doi.org/10.1016/j.physletb.2022.136908} {\bibfield  {journal}
  {\bibinfo  {journal} {Phys. Lett. B}\ }\textbf {\bibinfo {volume} {826}},\
  \bibinfo {pages} {136908} (\bibinfo {year} {2022})},\ \Eprint
  {https://arxiv.org/abs/2112.13218} {arXiv:2112.13218 [hep-ph]} \BibitemShut
  {NoStop}%
\bibitem [{\citenamefont {Qiao}\ and\ \citenamefont
  {Song}(2025)}]{Qiao:2024bgg}%
  \BibitemOpen
  \bibfield  {author} {\bibinfo {author} {\bibfnamefont {S.-Y.}\ \bibnamefont
  {Qiao}}\ and\ \bibinfo {author} {\bibfnamefont {Q.-T.}\ \bibnamefont
  {Song}},\ }\bibfield  {title} {\bibinfo {title} {{Twist-3 contribution in the
  Drell-Yan process with tensor-polarized deuteron}},\ }\href
  {https://doi.org/10.1103/PhysRevD.111.054026} {\bibfield  {journal} {\bibinfo
   {journal} {Phys. Rev. D}\ }\textbf {\bibinfo {volume} {111}},\ \bibinfo
  {pages} {054026} (\bibinfo {year} {2025})},\ \Eprint
  {https://arxiv.org/abs/2410.13225} {arXiv:2410.13225 [hep-ph]} \BibitemShut
  {NoStop}%
\bibitem [{\citenamefont {Cosyn}\ \emph {et~al.}(2025)\citenamefont {Cosyn},
  \citenamefont {Tomei}, \citenamefont {Sosa},\ and\ \citenamefont
  {Zec}}]{Cosyn:2024drt}%
  \BibitemOpen
  \bibfield  {author} {\bibinfo {author} {\bibfnamefont {W.}~\bibnamefont
  {Cosyn}}, \bibinfo {author} {\bibfnamefont {B.~R.}\ \bibnamefont {Tomei}},
  \bibinfo {author} {\bibfnamefont {A.}~\bibnamefont {Sosa}},\ and\ \bibinfo
  {author} {\bibfnamefont {A.}~\bibnamefont {Zec}},\ }\bibfield  {title}
  {\bibinfo {title} {{Polarization options in inclusive DIS off tensor
  polarized deuteron}},\ }\href
  {https://doi.org/10.1140/epja/s10050-025-01511-x} {\bibfield  {journal}
  {\bibinfo  {journal} {Eur. Phys. J. A}\ }\textbf {\bibinfo {volume} {61}},\
  \bibinfo {pages} {83} (\bibinfo {year} {2025})},\ \Eprint
  {https://arxiv.org/abs/2410.12764} {arXiv:2410.12764 [hep-ph]} \BibitemShut
  {NoStop}%
\bibitem [{\citenamefont {Kumano}\ and\ \citenamefont
  {Kuroki}(2026)}]{Kumano:2026xxv}%
  \BibitemOpen
  \bibfield  {author} {\bibinfo {author} {\bibfnamefont {S.}~\bibnamefont
  {Kumano}}\ and\ \bibinfo {author} {\bibfnamefont {K.}~\bibnamefont
  {Kuroki}},\ }\bibfield  {title} {\bibinfo {title} {{Twist-2 relations for the
  twist-3 tensor-polarized distribution function fLT of a spin-1 hadron by the
  operator-product-expansion method}},\ }\href
  {https://doi.org/10.1016/j.physletb.2026.140741} {\bibfield  {journal}
  {\bibinfo  {journal} {Phys. Lett. B}\ }\textbf {\bibinfo {volume} {880}},\
  \bibinfo {pages} {140741} (\bibinfo {year} {2026})},\ \Eprint
  {https://arxiv.org/abs/2605.00430} {arXiv:2605.00430 [hep-ph]} \BibitemShut
  {NoStop}%
\bibitem [{\citenamefont {Pascalutsa}\ and\ \citenamefont
  {Vanderhaeghen}(2026)}]{Pascalutsa:2026ubk}%
  \BibitemOpen
  \bibfield  {author} {\bibinfo {author} {\bibfnamefont {V.}~\bibnamefont
  {Pascalutsa}}\ and\ \bibinfo {author} {\bibfnamefont {M.}~\bibnamefont
  {Vanderhaeghen}},\ }\bibfield  {title} {\bibinfo {title} {{Tensor-structure
  sum rules for spin-1 targets at all $Q^2$}},\ }\href@noop {} {\  (\bibinfo
  {year} {2026})},\ \Eprint {https://arxiv.org/abs/2610.09862}
  {arXiv:2610.09862 [hep-ph]} \BibitemShut {NoStop}%
\bibitem [{\citenamefont {Bacchetta}\ and\ \citenamefont
  {Mulders}(2000)}]{Bacchetta:2000jk}%
  \BibitemOpen
  \bibfield  {author} {\bibinfo {author} {\bibfnamefont {A.}~\bibnamefont
  {Bacchetta}}\ and\ \bibinfo {author} {\bibfnamefont {P.~J.}\ \bibnamefont
  {Mulders}},\ }\bibfield  {title} {\bibinfo {title} {{Deep inelastic
  leptoproduction of spin-one hadrons}},\ }\href
  {https://doi.org/10.1103/PhysRevD.62.114004} {\bibfield  {journal} {\bibinfo
  {journal} {Phys. Rev. D}\ }\textbf {\bibinfo {volume} {62}},\ \bibinfo
  {pages} {114004} (\bibinfo {year} {2000})},\ \Eprint
  {https://arxiv.org/abs/hep-ph/0007120} {arXiv:hep-ph/0007120} \BibitemShut
  {NoStop}%
\bibitem [{\citenamefont {Bacchetta}\ and\ \citenamefont
  {Mulders}(2001)}]{Bacchetta:2001rb}%
  \BibitemOpen
  \bibfield  {author} {\bibinfo {author} {\bibfnamefont {A.}~\bibnamefont
  {Bacchetta}}\ and\ \bibinfo {author} {\bibfnamefont {P.~J.}\ \bibnamefont
  {Mulders}},\ }\bibfield  {title} {\bibinfo {title} {{Positivity bounds on
  spin one distribution and fragmentation functions}},\ }\href
  {https://doi.org/10.1016/S0370-2693(01)01051-6} {\bibfield  {journal}
  {\bibinfo  {journal} {Phys. Lett. B}\ }\textbf {\bibinfo {volume} {518}},\
  \bibinfo {pages} {85} (\bibinfo {year} {2001})},\ \Eprint
  {https://arxiv.org/abs/hep-ph/0104176} {arXiv:hep-ph/0104176} \BibitemShut
  {NoStop}%
\bibitem [{\citenamefont {Kumano}\ and\ \citenamefont
  {Song}(2021{\natexlab{b}})}]{Kumano:2020ijt}%
  \BibitemOpen
  \bibfield  {author} {\bibinfo {author} {\bibfnamefont {S.}~\bibnamefont
  {Kumano}}\ and\ \bibinfo {author} {\bibfnamefont {Q.-T.}\ \bibnamefont
  {Song}},\ }\bibfield  {title} {\bibinfo {title}
  {{Transverse-momentum-dependent parton distribution functions up to twist 4
  for spin-1 hadrons}},\ }\href {https://doi.org/10.1103/PhysRevD.103.014025}
  {\bibfield  {journal} {\bibinfo  {journal} {Phys. Rev. D}\ }\textbf {\bibinfo
  {volume} {103}},\ \bibinfo {pages} {014025} (\bibinfo {year}
  {2021}{\natexlab{b}})},\ \Eprint {https://arxiv.org/abs/2011.08583}
  {arXiv:2011.08583 [hep-ph]} \BibitemShut {NoStop}%
\bibitem [{\citenamefont {Ninomiya}\ \emph {et~al.}(2017)\citenamefont
  {Ninomiya}, \citenamefont {Bentz},\ and\ \citenamefont
  {Clo{\"e}t}}]{Ninomiya:2017ggn}%
  \BibitemOpen
  \bibfield  {author} {\bibinfo {author} {\bibfnamefont {Y.}~\bibnamefont
  {Ninomiya}}, \bibinfo {author} {\bibfnamefont {W.}~\bibnamefont {Bentz}},\
  and\ \bibinfo {author} {\bibfnamefont {I.~C.}\ \bibnamefont {Clo{\"e}t}},\
  }\bibfield  {title} {\bibinfo {title} {{Transverse-momentum-dependent quark
  distribution functions of spin-one targets: Formalism and covariant
  calculations}},\ }\href {https://doi.org/10.1103/PhysRevC.96.045206}
  {\bibfield  {journal} {\bibinfo  {journal} {Phys. Rev. C}\ }\textbf {\bibinfo
  {volume} {96}},\ \bibinfo {pages} {045206} (\bibinfo {year} {2017})},\
  \Eprint {https://arxiv.org/abs/1707.03787} {arXiv:1707.03787 [nucl-th]}
  \BibitemShut {NoStop}%
\bibitem [{\citenamefont {Cotogno}\ \emph {et~al.}(2017)\citenamefont
  {Cotogno}, \citenamefont {van Daal},\ and\ \citenamefont
  {Mulders}}]{Cotogno:2017puy}%
  \BibitemOpen
  \bibfield  {author} {\bibinfo {author} {\bibfnamefont {S.}~\bibnamefont
  {Cotogno}}, \bibinfo {author} {\bibfnamefont {T.}~\bibnamefont {van Daal}},\
  and\ \bibinfo {author} {\bibfnamefont {P.~J.}\ \bibnamefont {Mulders}},\
  }\bibfield  {title} {\bibinfo {title} {{Positivity bounds on gluon TMDs for
  hadrons of spin $\le$ 1}},\ }\href {https://doi.org/10.1007/JHEP11(2017)185}
  {\bibfield  {journal} {\bibinfo  {journal} {JHEP}\ }\textbf {\bibinfo
  {volume} {11}},\ \bibinfo {pages} {185}},\ \Eprint
  {https://arxiv.org/abs/1709.07827} {arXiv:1709.07827 [hep-ph]} \BibitemShut
  {NoStop}%
\bibitem [{\citenamefont {Zhao}\ \emph {et~al.}(2025)\citenamefont {Zhao},
  \citenamefont {Bacchetta}, \citenamefont {Kumano}, \citenamefont {Liu},\ and\
  \citenamefont {Zhou}}]{Zhao:2025vol}%
  \BibitemOpen
  \bibfield  {author} {\bibinfo {author} {\bibfnamefont {J.}~\bibnamefont
  {Zhao}}, \bibinfo {author} {\bibfnamefont {A.}~\bibnamefont {Bacchetta}},
  \bibinfo {author} {\bibfnamefont {S.}~\bibnamefont {Kumano}}, \bibinfo
  {author} {\bibfnamefont {T.}~\bibnamefont {Liu}},\ and\ \bibinfo {author}
  {\bibfnamefont {Y.-j.}\ \bibnamefont {Zhou}},\ }\bibfield  {title} {\bibinfo
  {title} {{Semi-inclusive deep inelastic scattering off a tensor-polarized
  spin-1 target}},\ }\href {https://doi.org/10.1007/JHEP12(2025)067} {\bibfield
   {journal} {\bibinfo  {journal} {JHEP}\ }\textbf {\bibinfo {volume} {12}},\
  \bibinfo {pages} {067}},\ \Eprint {https://arxiv.org/abs/2508.06134}
  {arXiv:2508.06134 [hep-ph]} \BibitemShut {NoStop}%
\bibitem [{\citenamefont {M{\"u}ller}\ \emph {et~al.}(1994)\citenamefont
  {M{\"u}ller}, \citenamefont {Robaschik}, \citenamefont {Geyer}, \citenamefont
  {Dittes},\ and\ \citenamefont {Ho{\v{r}}ej{\v{s}}i}}]{Muller:1994ses}%
  \BibitemOpen
  \bibfield  {author} {\bibinfo {author} {\bibfnamefont {D.}~\bibnamefont
  {M{\"u}ller}}, \bibinfo {author} {\bibfnamefont {D.}~\bibnamefont
  {Robaschik}}, \bibinfo {author} {\bibfnamefont {B.}~\bibnamefont {Geyer}},
  \bibinfo {author} {\bibfnamefont {F.~M.}\ \bibnamefont {Dittes}},\ and\
  \bibinfo {author} {\bibfnamefont {J.}~\bibnamefont {Ho{\v{r}}ej{\v{s}}i}},\
  }\bibfield  {title} {\bibinfo {title} {{Wave functions, evolution equations
  and evolution kernels from light ray operators of QCD}},\ }\href
  {https://doi.org/10.1002/prop.2190420202} {\bibfield  {journal} {\bibinfo
  {journal} {Fortsch. Phys.}\ }\textbf {\bibinfo {volume} {42}},\ \bibinfo
  {pages} {101} (\bibinfo {year} {1994})},\ \Eprint
  {https://arxiv.org/abs/hep-ph/9812448} {arXiv:hep-ph/9812448} \BibitemShut
  {NoStop}%
\bibitem [{\citenamefont {Ji}(1997)}]{Ji:1996ek}%
  \BibitemOpen
  \bibfield  {author} {\bibinfo {author} {\bibfnamefont {X.-D.}\ \bibnamefont
  {Ji}},\ }\bibfield  {title} {\bibinfo {title} {{Gauge-Invariant Decomposition
  of Nucleon Spin}},\ }\href {https://doi.org/10.1103/PhysRevLett.78.610}
  {\bibfield  {journal} {\bibinfo  {journal} {Phys. Rev. Lett.}\ }\textbf
  {\bibinfo {volume} {78}},\ \bibinfo {pages} {610} (\bibinfo {year} {1997})},\
  \Eprint {https://arxiv.org/abs/hep-ph/9603249} {arXiv:hep-ph/9603249}
  \BibitemShut {NoStop}%
\bibitem [{\citenamefont {Radyushkin}(1996)}]{Radyushkin:1996nd}%
  \BibitemOpen
  \bibfield  {author} {\bibinfo {author} {\bibfnamefont {A.~V.}\ \bibnamefont
  {Radyushkin}},\ }\bibfield  {title} {\bibinfo {title} {{Scaling limit of
  deeply virtual Compton scattering}},\ }\href
  {https://doi.org/10.1016/0370-2693(96)00528-X} {\bibfield  {journal}
  {\bibinfo  {journal} {Phys. Lett. B}\ }\textbf {\bibinfo {volume} {380}},\
  \bibinfo {pages} {417} (\bibinfo {year} {1996})},\ \Eprint
  {https://arxiv.org/abs/hep-ph/9604317} {arXiv:hep-ph/9604317} \BibitemShut
  {NoStop}%
\bibitem [{\citenamefont {Diehl}(2003)}]{Diehl:2003ny}%
  \BibitemOpen
  \bibfield  {author} {\bibinfo {author} {\bibfnamefont {M.}~\bibnamefont
  {Diehl}},\ }\bibfield  {title} {\bibinfo {title} {{Generalized parton
  distributions}},\ }\href {https://doi.org/10.1016/j.physrep.2003.08.002}
  {\bibfield  {journal} {\bibinfo  {journal} {Phys. Rept.}\ }\textbf {\bibinfo
  {volume} {388}},\ \bibinfo {pages} {41} (\bibinfo {year} {2003})},\ \Eprint
  {https://arxiv.org/abs/hep-ph/0307382} {arXiv:hep-ph/0307382} \BibitemShut
  {NoStop}%
\bibitem [{\citenamefont {Belitsky}\ and\ \citenamefont
  {Radyushkin}(2005)}]{Belitsky:2005qn}%
  \BibitemOpen
  \bibfield  {author} {\bibinfo {author} {\bibfnamefont {A.~V.}\ \bibnamefont
  {Belitsky}}\ and\ \bibinfo {author} {\bibfnamefont {A.~V.}\ \bibnamefont
  {Radyushkin}},\ }\bibfield  {title} {\bibinfo {title} {{Unraveling hadron
  structure with generalized parton distributions}},\ }\href
  {https://doi.org/10.1016/j.physrep.2005.06.002} {\bibfield  {journal}
  {\bibinfo  {journal} {Phys. Rept.}\ }\textbf {\bibinfo {volume} {418}},\
  \bibinfo {pages} {1} (\bibinfo {year} {2005})},\ \Eprint
  {https://arxiv.org/abs/hep-ph/0504030} {arXiv:hep-ph/0504030} \BibitemShut
  {NoStop}%
\bibitem [{\citenamefont {Guo}\ \emph {et~al.}(2026)\citenamefont {Guo},
  \citenamefont {Ji}, \citenamefont {Ji},\ and\ \citenamefont
  {Zhang}}]{Guo:2026swz}%
  \BibitemOpen
  \bibfield  {author} {\bibinfo {author} {\bibfnamefont {Y.}~\bibnamefont
  {Guo}}, \bibinfo {author} {\bibfnamefont {X.}~\bibnamefont {Ji}}, \bibinfo
  {author} {\bibfnamefont {Y.}~\bibnamefont {Ji}},\ and\ \bibinfo {author}
  {\bibfnamefont {J.}~\bibnamefont {Zhang}},\ }\bibfield  {title} {\bibinfo
  {title} {{Generalized parton distributions: Theory meets experiment}},\
  }\href@noop {} {\  (\bibinfo {year} {2026})},\ \Eprint
  {https://arxiv.org/abs/2609.19544} {arXiv:2609.19544 [hep-ph]} \BibitemShut
  {NoStop}%
\bibitem [{\citenamefont {Berger}\ \emph {et~al.}(2001)\citenamefont {Berger},
  \citenamefont {Cano}, \citenamefont {Diehl},\ and\ \citenamefont
  {Pire}}]{Berger:2001zb}%
  \BibitemOpen
  \bibfield  {author} {\bibinfo {author} {\bibfnamefont {E.~R.}\ \bibnamefont
  {Berger}}, \bibinfo {author} {\bibfnamefont {F.}~\bibnamefont {Cano}},
  \bibinfo {author} {\bibfnamefont {M.}~\bibnamefont {Diehl}},\ and\ \bibinfo
  {author} {\bibfnamefont {B.}~\bibnamefont {Pire}},\ }\bibfield  {title}
  {\bibinfo {title} {{Generalized parton distributions in the deuteron}},\
  }\href {https://doi.org/10.1103/PhysRevLett.87.142302} {\bibfield  {journal}
  {\bibinfo  {journal} {Phys. Rev. Lett.}\ }\textbf {\bibinfo {volume} {87}},\
  \bibinfo {pages} {142302} (\bibinfo {year} {2001})},\ \Eprint
  {https://arxiv.org/abs/hep-ph/0106192} {arXiv:hep-ph/0106192} \BibitemShut
  {NoStop}%
\bibitem [{\citenamefont {Dong}\ and\ \citenamefont
  {Liang}(2013)}]{Dong:2013rk}%
  \BibitemOpen
  \bibfield  {author} {\bibinfo {author} {\bibfnamefont {Y.}~\bibnamefont
  {Dong}}\ and\ \bibinfo {author} {\bibfnamefont {C.}~\bibnamefont {Liang}},\
  }\bibfield  {title} {\bibinfo {title} {{Generalized parton distribution
  functions of a deuteron in a phenomenological Lagrangian approach}},\ }\href
  {https://doi.org/10.1088/0954-3899/40/2/025001} {\bibfield  {journal}
  {\bibinfo  {journal} {J. Phys. G}\ }\textbf {\bibinfo {volume} {40}},\
  \bibinfo {pages} {025001} (\bibinfo {year} {2013})}\BibitemShut {NoStop}%
\bibitem [{\citenamefont {Cosyn}\ and\ \citenamefont
  {Pire}(2018)}]{Cosyn:2018rdm}%
  \BibitemOpen
  \bibfield  {author} {\bibinfo {author} {\bibfnamefont {W.}~\bibnamefont
  {Cosyn}}\ and\ \bibinfo {author} {\bibfnamefont {B.}~\bibnamefont {Pire}},\
  }\bibfield  {title} {\bibinfo {title} {{Transversity generalized parton
  distributions for the deuteron}},\ }\href
  {https://doi.org/10.1103/PhysRevD.98.074020} {\bibfield  {journal} {\bibinfo
  {journal} {Phys. Rev. D}\ }\textbf {\bibinfo {volume} {98}},\ \bibinfo
  {pages} {074020} (\bibinfo {year} {2018})},\ \Eprint
  {https://arxiv.org/abs/1806.01177} {arXiv:1806.01177 [hep-ph]} \BibitemShut
  {NoStop}%
\bibitem [{\citenamefont {Cosyn}\ \emph
  {et~al.}(2019{\natexlab{a}})\citenamefont {Cosyn}, \citenamefont {Freese},\
  and\ \citenamefont {Pire}}]{Cosyn:2018thq}%
  \BibitemOpen
  \bibfield  {author} {\bibinfo {author} {\bibfnamefont {W.}~\bibnamefont
  {Cosyn}}, \bibinfo {author} {\bibfnamefont {A.}~\bibnamefont {Freese}},\ and\
  \bibinfo {author} {\bibfnamefont {B.}~\bibnamefont {Pire}},\ }\bibfield
  {title} {\bibinfo {title} {{Polynomiality sum rules for generalized parton
  distributions of spin-1 targets}},\ }\href
  {https://doi.org/10.1103/PhysRevD.99.094035} {\bibfield  {journal} {\bibinfo
  {journal} {Phys. Rev. D}\ }\textbf {\bibinfo {volume} {99}},\ \bibinfo
  {pages} {094035} (\bibinfo {year} {2019}{\natexlab{a}})},\ \Eprint
  {https://arxiv.org/abs/1812.01511} {arXiv:1812.01511 [hep-ph]} \BibitemShut
  {NoStop}%
\bibitem [{\citenamefont {Polyakov}\ and\ \citenamefont
  {Schweitzer}(2018)}]{Polyakov:2018zvc}%
  \BibitemOpen
  \bibfield  {author} {\bibinfo {author} {\bibfnamefont {M.~V.}\ \bibnamefont
  {Polyakov}}\ and\ \bibinfo {author} {\bibfnamefont {P.}~\bibnamefont
  {Schweitzer}},\ }\bibfield  {title} {\bibinfo {title} {{Forces inside
  hadrons: pressure, surface tension, mechanical radius, and all that}},\
  }\href {https://doi.org/10.1142/S0217751X18300259} {\bibfield  {journal}
  {\bibinfo  {journal} {Int. J. Mod. Phys. A}\ }\textbf {\bibinfo {volume}
  {33}},\ \bibinfo {pages} {1830025} (\bibinfo {year} {2018})},\ \Eprint
  {https://arxiv.org/abs/1805.06596} {arXiv:1805.06596 [hep-ph]} \BibitemShut
  {NoStop}%
\bibitem [{\citenamefont {Kumano}\ \emph {et~al.}(2018)\citenamefont {Kumano},
  \citenamefont {Song},\ and\ \citenamefont {Teryaev}}]{Kumano:2017lhr}%
  \BibitemOpen
  \bibfield  {author} {\bibinfo {author} {\bibfnamefont {S.}~\bibnamefont
  {Kumano}}, \bibinfo {author} {\bibfnamefont {Q.-T.}\ \bibnamefont {Song}},\
  and\ \bibinfo {author} {\bibfnamefont {O.~V.}\ \bibnamefont {Teryaev}},\
  }\bibfield  {title} {\bibinfo {title} {{Hadron tomography by generalized
  distribution amplitudes in pion-pair production process $\gamma^* \gamma
  \rightarrow \pi^0 \pi^0 $ and gravitational form factors for pion}},\ }\href
  {https://doi.org/10.1103/PhysRevD.97.014020} {\bibfield  {journal} {\bibinfo
  {journal} {Phys. Rev. D}\ }\textbf {\bibinfo {volume} {97}},\ \bibinfo
  {pages} {014020} (\bibinfo {year} {2018})},\ \Eprint
  {https://arxiv.org/abs/1711.08088} {arXiv:1711.08088 [hep-ph]} \BibitemShut
  {NoStop}%
\bibitem [{\citenamefont {Taneja}\ \emph {et~al.}(2012)\citenamefont {Taneja},
  \citenamefont {Kathuria}, \citenamefont {Liuti},\ and\ \citenamefont
  {Goldstein}}]{Taneja:2011sy}%
  \BibitemOpen
  \bibfield  {author} {\bibinfo {author} {\bibfnamefont {S.~K.}\ \bibnamefont
  {Taneja}}, \bibinfo {author} {\bibfnamefont {K.}~\bibnamefont {Kathuria}},
  \bibinfo {author} {\bibfnamefont {S.}~\bibnamefont {Liuti}},\ and\ \bibinfo
  {author} {\bibfnamefont {G.~R.}\ \bibnamefont {Goldstein}},\ }\bibfield
  {title} {\bibinfo {title} {{Angular momentum sum rule for spin one hadronic
  systems}},\ }\href {https://doi.org/10.1103/PhysRevD.86.036008} {\bibfield
  {journal} {\bibinfo  {journal} {Phys. Rev. D}\ }\textbf {\bibinfo {volume}
  {86}},\ \bibinfo {pages} {036008} (\bibinfo {year} {2012})},\ \Eprint
  {https://arxiv.org/abs/1101.0581} {arXiv:1101.0581 [hep-ph]} \BibitemShut
  {NoStop}%
\bibitem [{\citenamefont {Burkert}\ \emph {et~al.}(2023)\citenamefont
  {Burkert}, \citenamefont {Elouadrhiri}, \citenamefont {Girod}, \citenamefont
  {Lorc\'e}, \citenamefont {Schweitzer},\ and\ \citenamefont
  {Shanahan}}]{Burkert:2023wzr}%
  \BibitemOpen
  \bibfield  {author} {\bibinfo {author} {\bibfnamefont {V.~D.}\ \bibnamefont
  {Burkert}}, \bibinfo {author} {\bibfnamefont {L.}~\bibnamefont
  {Elouadrhiri}}, \bibinfo {author} {\bibfnamefont {F.~X.}\ \bibnamefont
  {Girod}}, \bibinfo {author} {\bibfnamefont {C.}~\bibnamefont {Lorc\'e}},
  \bibinfo {author} {\bibfnamefont {P.}~\bibnamefont {Schweitzer}},\ and\
  \bibinfo {author} {\bibfnamefont {P.~E.}\ \bibnamefont {Shanahan}},\
  }\bibfield  {title} {\bibinfo {title} {{Colloquium: Gravitational form
  factors of the proton}},\ }\href
  {https://doi.org/10.1103/RevModPhys.95.041002} {\bibfield  {journal}
  {\bibinfo  {journal} {Rev. Mod. Phys.}\ }\textbf {\bibinfo {volume} {95}},\
  \bibinfo {pages} {041002} (\bibinfo {year} {2023})},\ \Eprint
  {https://arxiv.org/abs/2303.08347} {arXiv:2303.08347 [hep-ph]} \BibitemShut
  {NoStop}%
\bibitem [{\citenamefont {Freese}\ and\ \citenamefont
  {Miller}(2021)}]{Freese:2021qtb}%
  \BibitemOpen
  \bibfield  {author} {\bibinfo {author} {\bibfnamefont {A.}~\bibnamefont
  {Freese}}\ and\ \bibinfo {author} {\bibfnamefont {G.~A.}\ \bibnamefont
  {Miller}},\ }\bibfield  {title} {\bibinfo {title} {{Genuine empirical
  pressure within the proton}},\ }\href
  {https://doi.org/10.1103/PhysRevD.104.014024} {\bibfield  {journal} {\bibinfo
   {journal} {Phys. Rev. D}\ }\textbf {\bibinfo {volume} {104}},\ \bibinfo
  {pages} {014024} (\bibinfo {year} {2021})},\ \Eprint
  {https://arxiv.org/abs/2104.03213} {arXiv:2104.03213 [hep-ph]} \BibitemShut
  {NoStop}%
\bibitem [{\citenamefont {Lorc\'e}\ and\ \citenamefont
  {Song}(2025)}]{Lorce:2025ayr}%
  \BibitemOpen
  \bibfield  {author} {\bibinfo {author} {\bibfnamefont {C.}~\bibnamefont
  {Lorc\'e}}\ and\ \bibinfo {author} {\bibfnamefont {Q.-T.}\ \bibnamefont
  {Song}},\ }\bibfield  {title} {\bibinfo {title} {{Spin-orbit correlation and
  spatial distributions for spin-0 hadrons}},\ }\href
  {https://doi.org/10.1016/j.physletb.2025.139433} {\bibfield  {journal}
  {\bibinfo  {journal} {Phys. Lett. B}\ }\textbf {\bibinfo {volume} {864}},\
  \bibinfo {pages} {139433} (\bibinfo {year} {2025})},\ \Eprint
  {https://arxiv.org/abs/2501.05092} {arXiv:2501.05092 [hep-ph]} \BibitemShut
  {NoStop}%
\bibitem [{\citenamefont {Fujii}\ and\ \citenamefont
  {Tanaka}(2025)}]{Fujii:2025pkv}%
  \BibitemOpen
  \bibfield  {author} {\bibinfo {author} {\bibfnamefont {D.}~\bibnamefont
  {Fujii}}\ and\ \bibinfo {author} {\bibfnamefont {M.}~\bibnamefont {Tanaka}},\
  }\bibfield  {title} {\bibinfo {title} {{Scale-anomaly-induced binding
  pressure in hadrons}},\ }\href
  {https://doi.org/10.1016/j.physletb.2025.139872} {\bibfield  {journal}
  {\bibinfo  {journal} {Phys. Lett. B}\ }\textbf {\bibinfo {volume} {870}},\
  \bibinfo {pages} {139872} (\bibinfo {year} {2025})}\BibitemShut {NoStop}%
\bibitem [{\citenamefont {Polyakov}\ and\ \citenamefont
  {Sun}(2019)}]{Polyakov:2019lbq}%
  \BibitemOpen
  \bibfield  {author} {\bibinfo {author} {\bibfnamefont {M.~V.}\ \bibnamefont
  {Polyakov}}\ and\ \bibinfo {author} {\bibfnamefont {B.-D.}\ \bibnamefont
  {Sun}},\ }\bibfield  {title} {\bibinfo {title} {{Gravitational form factors
  of a spin one particle}},\ }\href
  {https://doi.org/10.1103/PhysRevD.100.036003} {\bibfield  {journal} {\bibinfo
   {journal} {Phys. Rev. D}\ }\textbf {\bibinfo {volume} {100}},\ \bibinfo
  {pages} {036003} (\bibinfo {year} {2019})},\ \Eprint
  {https://arxiv.org/abs/1903.02738} {arXiv:1903.02738 [hep-ph]} \BibitemShut
  {NoStop}%
\bibitem [{\citenamefont {Cosyn}\ \emph
  {et~al.}(2019{\natexlab{b}})\citenamefont {Cosyn}, \citenamefont {Cotogno},
  \citenamefont {Freese},\ and\ \citenamefont {Lorc{\'e}}}]{Cosyn:2019aio}%
  \BibitemOpen
  \bibfield  {author} {\bibinfo {author} {\bibfnamefont {W.}~\bibnamefont
  {Cosyn}}, \bibinfo {author} {\bibfnamefont {S.}~\bibnamefont {Cotogno}},
  \bibinfo {author} {\bibfnamefont {A.}~\bibnamefont {Freese}},\ and\ \bibinfo
  {author} {\bibfnamefont {C.}~\bibnamefont {Lorc{\'e}}},\ }\bibfield  {title}
  {\bibinfo {title} {{The energy-momentum tensor of spin-1 hadrons:
  formalism}},\ }\href {https://doi.org/10.1140/epjc/s10052-019-6981-3}
  {\bibfield  {journal} {\bibinfo  {journal} {Eur. Phys. J. C}\ }\textbf
  {\bibinfo {volume} {79}},\ \bibinfo {pages} {476} (\bibinfo {year}
  {2019}{\natexlab{b}})},\ \Eprint {https://arxiv.org/abs/1903.00408}
  {arXiv:1903.00408 [hep-ph]} \BibitemShut {NoStop}%
\bibitem [{\citenamefont {Benali}\ \emph {et~al.}(2020)\citenamefont {Benali}
  \emph {et~al.}}]{Benali:2020vma}%
  \BibitemOpen
  \bibfield  {author} {\bibinfo {author} {\bibfnamefont {M.}~\bibnamefont
  {Benali}} \emph {et~al.},\ }\bibfield  {title} {\bibinfo {title} {{Deeply
  virtual Compton scattering off the neutron}},\ }\href
  {https://doi.org/10.1038/s41567-019-0774-3} {\bibfield  {journal} {\bibinfo
  {journal} {Nature Phys.}\ }\textbf {\bibinfo {volume} {16}},\ \bibinfo
  {pages} {191} (\bibinfo {year} {2020})},\ \Eprint
  {https://arxiv.org/abs/2109.02076} {arXiv:2109.02076 [hep-ph]} \BibitemShut
  {NoStop}%
\bibitem [{\citenamefont {Cano}\ and\ \citenamefont
  {Pire}(2002)}]{Cano:2002ph}%
  \BibitemOpen
  \bibfield  {author} {\bibinfo {author} {\bibfnamefont {F.}~\bibnamefont
  {Cano}}\ and\ \bibinfo {author} {\bibfnamefont {B.}~\bibnamefont {Pire}},\
  }\bibfield  {title} {\bibinfo {title} {{Deeply virtual Compton scattering on
  the deuteron}},\ }\href {https://doi.org/10.1016/S0375-9474(02)01204-6}
  {\bibfield  {journal} {\bibinfo  {journal} {Nucl. Phys. A}\ }\textbf
  {\bibinfo {volume} {711}},\ \bibinfo {pages} {133} (\bibinfo {year}
  {2002})},\ \Eprint {https://arxiv.org/abs/hep-ph/0206215}
  {arXiv:hep-ph/0206215} \BibitemShut {NoStop}%
\bibitem [{\citenamefont {Kirchner}\ and\ \citenamefont
  {Mueller}(2003{\natexlab{a}})}]{Kirchner:2002zu}%
  \BibitemOpen
  \bibfield  {author} {\bibinfo {author} {\bibfnamefont {A.}~\bibnamefont
  {Kirchner}}\ and\ \bibinfo {author} {\bibfnamefont {D.}~\bibnamefont
  {Mueller}},\ }\bibfield  {title} {\bibinfo {title} {{Predictions for deeply
  virtual Compton scattering on a spin one target}},\ }\href@noop {} {\bibfield
   {journal} {\bibinfo  {journal} {Lect. Notes Phys.}\ }\textbf {\bibinfo
  {volume} {616}},\ \bibinfo {pages} {448} (\bibinfo {year}
  {2003}{\natexlab{a}})},\ \Eprint {https://arxiv.org/abs/hep-ph/0202279}
  {arXiv:hep-ph/0202279} \BibitemShut {NoStop}%
\bibitem [{\citenamefont {Kirchner}\ and\ \citenamefont
  {Mueller}(2003{\natexlab{b}})}]{Kirchner:2003wt}%
  \BibitemOpen
  \bibfield  {author} {\bibinfo {author} {\bibfnamefont {A.}~\bibnamefont
  {Kirchner}}\ and\ \bibinfo {author} {\bibfnamefont {D.}~\bibnamefont
  {Mueller}},\ }\bibfield  {title} {\bibinfo {title} {{Deeply virtual Compton
  scattering off nuclei}},\ }\href {https://doi.org/10.1140/epjc/s2003-01415-x}
  {\bibfield  {journal} {\bibinfo  {journal} {Eur. Phys. J. C}\ }\textbf
  {\bibinfo {volume} {32}},\ \bibinfo {pages} {347} (\bibinfo {year}
  {2003}{\natexlab{b}})},\ \Eprint {https://arxiv.org/abs/hep-ph/0302007}
  {arXiv:hep-ph/0302007} \BibitemShut {NoStop}%
\bibitem [{\citenamefont {Anikin}\ \emph {et~al.}(2010)\citenamefont {Anikin},
  \citenamefont {Pasechnik}, \citenamefont {Pire},\ and\ \citenamefont
  {Teryaev}}]{Anikin:2010ng}%
  \BibitemOpen
  \bibfield  {author} {\bibinfo {author} {\bibfnamefont {I.~V.}\ \bibnamefont
  {Anikin}}, \bibinfo {author} {\bibfnamefont {R.~S.}\ \bibnamefont
  {Pasechnik}}, \bibinfo {author} {\bibfnamefont {B.}~\bibnamefont {Pire}},\
  and\ \bibinfo {author} {\bibfnamefont {O.~V.}\ \bibnamefont {Teryaev}},\
  }\bibfield  {title} {\bibinfo {title} {{DVCS off deuteron and twist three
  contributions}},\ }\href {https://doi.org/10.22323/1.120.0137} {\bibfield
  {journal} {\bibinfo  {journal} {PoS}\ }\textbf {\bibinfo {volume}
  {ICHEP2010}},\ \bibinfo {pages} {137} (\bibinfo {year} {2010})},\ \Eprint
  {https://arxiv.org/abs/1011.2853} {arXiv:1011.2853 [hep-ph]} \BibitemShut
  {NoStop}%
\bibitem [{\citenamefont {Anikin}\ \emph {et~al.}(2012)\citenamefont {Anikin},
  \citenamefont {Pasechnik}, \citenamefont {Pire},\ and\ \citenamefont
  {Teryaev}}]{Anikin:2011aa}%
  \BibitemOpen
  \bibfield  {author} {\bibinfo {author} {\bibfnamefont {I.~V.}\ \bibnamefont
  {Anikin}}, \bibinfo {author} {\bibfnamefont {R.~S.}\ \bibnamefont
  {Pasechnik}}, \bibinfo {author} {\bibfnamefont {B.}~\bibnamefont {Pire}},\
  and\ \bibinfo {author} {\bibfnamefont {O.~V.}\ \bibnamefont {Teryaev}},\
  }\bibfield  {title} {\bibinfo {title} {{Gauge Invariance of DVCS off an
  Arbitrary Spin Hadron: The Deuteron Target Case}},\ }\href
  {https://doi.org/10.1140/epjc/s10052-012-2055-5} {\bibfield  {journal}
  {\bibinfo  {journal} {Eur. Phys. J. C}\ }\textbf {\bibinfo {volume} {72}},\
  \bibinfo {pages} {2055} (\bibinfo {year} {2012})},\ \Eprint
  {https://arxiv.org/abs/1112.1849} {arXiv:1112.1849 [hep-ph]} \BibitemShut
  {NoStop}%
\bibitem [{\citenamefont {Cano}\ and\ \citenamefont
  {Pire}(2004)}]{Cano:2003ju}%
  \BibitemOpen
  \bibfield  {author} {\bibinfo {author} {\bibfnamefont {F.}~\bibnamefont
  {Cano}}\ and\ \bibinfo {author} {\bibfnamefont {B.}~\bibnamefont {Pire}},\
  }\bibfield  {title} {\bibinfo {title} {{Deep electroproduction of photons and
  mesons on the deuteron}},\ }\href
  {https://doi.org/10.1140/epja/i2003-10127-x} {\bibfield  {journal} {\bibinfo
  {journal} {Eur. Phys. J. A}\ }\textbf {\bibinfo {volume} {19}},\ \bibinfo
  {pages} {423} (\bibinfo {year} {2004})},\ \Eprint
  {https://arxiv.org/abs/hep-ph/0307231} {arXiv:hep-ph/0307231} \BibitemShut
  {NoStop}%
\bibitem [{\citenamefont {Mazouz}\ \emph {et~al.}(2017)\citenamefont {Mazouz}
  \emph {et~al.}}]{JeffersonLabHallA:2017hky}%
  \BibitemOpen
  \bibfield  {author} {\bibinfo {author} {\bibfnamefont {M.}~\bibnamefont
  {Mazouz}} \emph {et~al.} (\bibinfo {collaboration} {Jefferson Lab Hall A}),\
  }\bibfield  {title} {\bibinfo {title} {{Rosenbluth separation of the $\pi^0$
  Electroproduction Cross Section off the Neutron}},\ }\href
  {https://doi.org/10.1103/PhysRevLett.118.222002} {\bibfield  {journal}
  {\bibinfo  {journal} {Phys. Rev. Lett.}\ }\textbf {\bibinfo {volume} {118}},\
  \bibinfo {pages} {222002} (\bibinfo {year} {2017})},\ \Eprint
  {https://arxiv.org/abs/1702.00835} {arXiv:1702.00835 [hep-ex]} \BibitemShut
  {NoStop}%
\bibitem [{\citenamefont {Berger}\ \emph {et~al.}(2002)\citenamefont {Berger},
  \citenamefont {Diehl},\ and\ \citenamefont {Pire}}]{Berger:2001xd}%
  \BibitemOpen
  \bibfield  {author} {\bibinfo {author} {\bibfnamefont {E.~R.}\ \bibnamefont
  {Berger}}, \bibinfo {author} {\bibfnamefont {M.}~\bibnamefont {Diehl}},\ and\
  \bibinfo {author} {\bibfnamefont {B.}~\bibnamefont {Pire}},\ }\bibfield
  {title} {\bibinfo {title} {{Time - like Compton scattering: Exclusive
  photoproduction of lepton pairs}},\ }\href
  {https://doi.org/10.1007/s100520200917} {\bibfield  {journal} {\bibinfo
  {journal} {Eur. Phys. J. C}\ }\textbf {\bibinfo {volume} {23}},\ \bibinfo
  {pages} {675} (\bibinfo {year} {2002})},\ \Eprint
  {https://arxiv.org/abs/hep-ph/0110062} {arXiv:hep-ph/0110062} \BibitemShut
  {NoStop}%
\bibitem [{\citenamefont {Mueller}\ \emph {et~al.}(2012)\citenamefont
  {Mueller}, \citenamefont {Pire}, \citenamefont {Szymanowski},\ and\
  \citenamefont {Wagner}}]{Mueller:2012sma}%
  \BibitemOpen
  \bibfield  {author} {\bibinfo {author} {\bibfnamefont {D.}~\bibnamefont
  {Mueller}}, \bibinfo {author} {\bibfnamefont {B.}~\bibnamefont {Pire}},
  \bibinfo {author} {\bibfnamefont {L.}~\bibnamefont {Szymanowski}},\ and\
  \bibinfo {author} {\bibfnamefont {J.}~\bibnamefont {Wagner}},\ }\bibfield
  {title} {\bibinfo {title} {{On timelike and spacelike hard exclusive
  reactions}},\ }\href {https://doi.org/10.1103/PhysRevD.86.031502} {\bibfield
  {journal} {\bibinfo  {journal} {Phys. Rev. D}\ }\textbf {\bibinfo {volume}
  {86}},\ \bibinfo {pages} {031502} (\bibinfo {year} {2012})},\ \Eprint
  {https://arxiv.org/abs/1203.4392} {arXiv:1203.4392 [hep-ph]} \BibitemShut
  {NoStop}%
\bibitem [{\citenamefont {Grocholski}\ \emph {et~al.}(2020)\citenamefont
  {Grocholski}, \citenamefont {Moutarde}, \citenamefont {Pire}, \citenamefont
  {Sznajder},\ and\ \citenamefont {Wagner}}]{Grocholski:2019pqj}%
  \BibitemOpen
  \bibfield  {author} {\bibinfo {author} {\bibfnamefont {O.}~\bibnamefont
  {Grocholski}}, \bibinfo {author} {\bibfnamefont {H.}~\bibnamefont
  {Moutarde}}, \bibinfo {author} {\bibfnamefont {B.}~\bibnamefont {Pire}},
  \bibinfo {author} {\bibfnamefont {P.}~\bibnamefont {Sznajder}},\ and\
  \bibinfo {author} {\bibfnamefont {J.}~\bibnamefont {Wagner}},\ }\bibfield
  {title} {\bibinfo {title} {{Data-driven study of timelike Compton
  scattering}},\ }\href {https://doi.org/10.1140/epjc/s10052-020-7700-9}
  {\bibfield  {journal} {\bibinfo  {journal} {Eur. Phys. J. C}\ }\textbf
  {\bibinfo {volume} {80}},\ \bibinfo {pages} {171} (\bibinfo {year} {2020})},\
  \Eprint {https://arxiv.org/abs/1912.09853} {arXiv:1912.09853 [hep-ph]}
  \BibitemShut {NoStop}%
\bibitem [{\citenamefont {Chatagnon}\ \emph {et~al.}(2021)\citenamefont
  {Chatagnon} \emph {et~al.}}]{CLAS:2021lky}%
  \BibitemOpen
  \bibfield  {author} {\bibinfo {author} {\bibfnamefont {P.}~\bibnamefont
  {Chatagnon}} \emph {et~al.} (\bibinfo {collaboration} {CLAS}),\ }\bibfield
  {title} {\bibinfo {title} {{First Measurement of Timelike Compton
  Scattering}},\ }\href {https://doi.org/10.1103/PhysRevLett.127.262501}
  {\bibfield  {journal} {\bibinfo  {journal} {Phys. Rev. Lett.}\ }\textbf
  {\bibinfo {volume} {127}},\ \bibinfo {pages} {262501} (\bibinfo {year}
  {2021})},\ \Eprint {https://arxiv.org/abs/2108.11746} {arXiv:2108.11746
  [hep-ex]} \BibitemShut {NoStop}%
\bibitem [{\citenamefont {Alvarado}\ \emph {et~al.}(2026)\citenamefont
  {Alvarado} \emph {et~al.}}]{Alvarado:2026ggy}%
  \BibitemOpen
  \bibfield  {author} {\bibinfo {author} {\bibfnamefont {J.~S.}\ \bibnamefont
  {Alvarado}} \emph {et~al.},\ }\bibfield  {title} {\bibinfo {title} {{Electro-
  and photoproduction of muon pairs with $\mu $CLAS12: double deeply virtual
  compton scattering, timelike compton scattering, and $J/\psi $ production}},\
  }\href {https://doi.org/10.1140/epja/s10050-026-01909-1} {\bibfield
  {journal} {\bibinfo  {journal} {Eur. Phys. J. A}\ }\textbf {\bibinfo {volume}
  {62}},\ \bibinfo {pages} {157} (\bibinfo {year} {2026})},\ \Eprint
  {https://arxiv.org/abs/2605.11690} {arXiv:2605.11690 [hep-ex]} \BibitemShut
  {NoStop}%
\bibitem [{\citenamefont {Cai}\ \emph {et~al.}(2024)\citenamefont {Cai},
  \citenamefont {Liu},\ and\ \citenamefont {Ma}}]{Cai:2022qza}%
  \BibitemOpen
  \bibfield  {author} {\bibinfo {author} {\bibfnamefont {M.}~\bibnamefont
  {Cai}}, \bibinfo {author} {\bibfnamefont {T.}~\bibnamefont {Liu}},\ and\
  \bibinfo {author} {\bibfnamefont {B.-Q.}\ \bibnamefont {Ma}},\ }\bibfield
  {title} {\bibinfo {title} {{Bethe-Heitler lepton pair production in the
  deuteron breakup reaction*}},\ }\href
  {https://doi.org/10.1088/1674-1137/ad0375} {\bibfield  {journal} {\bibinfo
  {journal} {Chin. Phys. C}\ }\textbf {\bibinfo {volume} {48}},\ \bibinfo
  {pages} {014103} (\bibinfo {year} {2024})},\ \Eprint
  {https://arxiv.org/abs/2210.10560} {arXiv:2210.10560 [hep-ph]} \BibitemShut
  {NoStop}%
\bibitem [{\citenamefont {Braun}\ and\ \citenamefont
  {Manashov}(2012)}]{Braun:2011dg}%
  \BibitemOpen
  \bibfield  {author} {\bibinfo {author} {\bibfnamefont {V.~M.}\ \bibnamefont
  {Braun}}\ and\ \bibinfo {author} {\bibfnamefont {A.~N.}\ \bibnamefont
  {Manashov}},\ }\bibfield  {title} {\bibinfo {title} {{Operator product
  expansion in QCD in off-forward kinematics: Separation of kinematic and
  dynamical contributions}},\ }\href {https://doi.org/10.1007/JHEP01(2012)085}
  {\bibfield  {journal} {\bibinfo  {journal} {JHEP}\ }\textbf {\bibinfo
  {volume} {01}},\ \bibinfo {pages} {085}},\ \Eprint
  {https://arxiv.org/abs/1111.6765} {arXiv:1111.6765 [hep-ph]} \BibitemShut
  {NoStop}%
\bibitem [{\citenamefont {Braun}\ and\ \citenamefont
  {Manashov}(2011)}]{Braun:2011zr}%
  \BibitemOpen
  \bibfield  {author} {\bibinfo {author} {\bibfnamefont {V.~M.}\ \bibnamefont
  {Braun}}\ and\ \bibinfo {author} {\bibfnamefont {A.~N.}\ \bibnamefont
  {Manashov}},\ }\bibfield  {title} {\bibinfo {title} {{Kinematic power
  corrections in off-forward hard reactions}},\ }\href
  {https://doi.org/10.1103/PhysRevLett.107.202001} {\bibfield  {journal}
  {\bibinfo  {journal} {Phys. Rev. Lett.}\ }\textbf {\bibinfo {volume} {107}},\
  \bibinfo {pages} {202001} (\bibinfo {year} {2011})},\ \Eprint
  {https://arxiv.org/abs/1108.2394} {arXiv:1108.2394 [hep-ph]} \BibitemShut
  {NoStop}%
\bibitem [{\citenamefont {Martinez-Fernandez}\ \emph
  {et~al.}(2025)\citenamefont {Martinez-Fernandez}, \citenamefont {Pire},
  \citenamefont {Sznajder},\ and\ \citenamefont
  {Wagner}}]{Martinez-Fernandez:2025gub}%
  \BibitemOpen
  \bibfield  {author} {\bibinfo {author} {\bibfnamefont {V.}~\bibnamefont
  {Martinez-Fernandez}}, \bibinfo {author} {\bibfnamefont {B.}~\bibnamefont
  {Pire}}, \bibinfo {author} {\bibfnamefont {P.}~\bibnamefont {Sznajder}},\
  and\ \bibinfo {author} {\bibfnamefont {J.}~\bibnamefont {Wagner}},\
  }\bibfield  {title} {\bibinfo {title} {{Timelike Compton scattering on a
  spin-0 target with kinematic twist-4 precision}},\ }\href
  {https://doi.org/10.1103/PhysRevD.111.074034} {\bibfield  {journal} {\bibinfo
   {journal} {Phys. Rev. D}\ }\textbf {\bibinfo {volume} {111}},\ \bibinfo
  {pages} {074034} (\bibinfo {year} {2025})},\ \Eprint
  {https://arxiv.org/abs/2503.02461} {arXiv:2503.02461 [hep-ph]} \BibitemShut
  {NoStop}%
\bibitem [{\citenamefont {Mart{\'\i}nez-Fern{\'a}ndez}\ \emph
  {et~al.}(2026)\citenamefont {Mart{\'\i}nez-Fern{\'a}ndez}, \citenamefont
  {Pire}, \citenamefont {Sznajder},\ and\ \citenamefont
  {Wagner}}]{Martinez-Fernandez:2026web}%
  \BibitemOpen
  \bibfield  {author} {\bibinfo {author} {\bibfnamefont {V.}~\bibnamefont
  {Mart{\'\i}nez-Fern{\'a}ndez}}, \bibinfo {author} {\bibfnamefont
  {B.}~\bibnamefont {Pire}}, \bibinfo {author} {\bibfnamefont {P.}~\bibnamefont
  {Sznajder}},\ and\ \bibinfo {author} {\bibfnamefont {J.}~\bibnamefont
  {Wagner}},\ }\bibfield  {title} {\bibinfo {title} {{Coherent deeply virtual
  Compton scattering on He4 beyond leading power}},\ }\href
  {https://doi.org/10.1103/h7kn-ql7j} {\bibfield  {journal} {\bibinfo
  {journal} {Phys. Rev. C}\ }\textbf {\bibinfo {volume} {114}},\ \bibinfo
  {pages} {035203} (\bibinfo {year} {2026})},\ \Eprint
  {https://arxiv.org/abs/2604.25677} {arXiv:2604.25677 [hep-ph]} \BibitemShut
  {NoStop}%
\bibitem [{\citenamefont {Moutarde}\ \emph {et~al.}(2013)\citenamefont
  {Moutarde}, \citenamefont {Pire}, \citenamefont {Sabatie}, \citenamefont
  {Szymanowski},\ and\ \citenamefont {Wagner}}]{Moutarde:2013qs}%
  \BibitemOpen
  \bibfield  {author} {\bibinfo {author} {\bibfnamefont {H.}~\bibnamefont
  {Moutarde}}, \bibinfo {author} {\bibfnamefont {B.}~\bibnamefont {Pire}},
  \bibinfo {author} {\bibfnamefont {F.}~\bibnamefont {Sabatie}}, \bibinfo
  {author} {\bibfnamefont {L.}~\bibnamefont {Szymanowski}},\ and\ \bibinfo
  {author} {\bibfnamefont {J.}~\bibnamefont {Wagner}},\ }\bibfield  {title}
  {\bibinfo {title} {{Timelike and spacelike deeply virtual Compton scattering
  at next-to-leading order}},\ }\href
  {https://doi.org/10.1103/PhysRevD.87.054029} {\bibfield  {journal} {\bibinfo
  {journal} {Phys. Rev. D}\ }\textbf {\bibinfo {volume} {87}},\ \bibinfo
  {pages} {054029} (\bibinfo {year} {2013})},\ \Eprint
  {https://arxiv.org/abs/1301.3819} {arXiv:1301.3819 [hep-ph]} \BibitemShut
  {NoStop}%
\bibitem [{\citenamefont {Gutsche}\ \emph {et~al.}(2015)\citenamefont
  {Gutsche}, \citenamefont {Lyubovitskij}, \citenamefont {Schmidt},\ and\
  \citenamefont {Vega}}]{Gutsche:2015xva}%
  \BibitemOpen
  \bibfield  {author} {\bibinfo {author} {\bibfnamefont {T.}~\bibnamefont
  {Gutsche}}, \bibinfo {author} {\bibfnamefont {V.~E.}\ \bibnamefont
  {Lyubovitskij}}, \bibinfo {author} {\bibfnamefont {I.}~\bibnamefont
  {Schmidt}},\ and\ \bibinfo {author} {\bibfnamefont {A.}~\bibnamefont
  {Vega}},\ }\bibfield  {title} {\bibinfo {title} {{Nuclear physics in
  soft-wall AdS/QCD: Deuteron electromagnetic form factors}},\ }\href
  {https://doi.org/10.1103/PhysRevD.91.114001} {\bibfield  {journal} {\bibinfo
  {journal} {Phys. Rev. D}\ }\textbf {\bibinfo {volume} {91}},\ \bibinfo
  {pages} {114001} (\bibinfo {year} {2015})},\ \Eprint
  {https://arxiv.org/abs/1501.02738} {arXiv:1501.02738 [hep-ph]} \BibitemShut
  {NoStop}%
\bibitem [{\citenamefont {Heller}\ \emph {et~al.}(2021)\citenamefont {Heller},
  \citenamefont {Keil},\ and\ \citenamefont {Vanderhaeghen}}]{Heller:2020lnm}%
  \BibitemOpen
  \bibfield  {author} {\bibinfo {author} {\bibfnamefont {M.}~\bibnamefont
  {Heller}}, \bibinfo {author} {\bibfnamefont {N.}~\bibnamefont {Keil}},\ and\
  \bibinfo {author} {\bibfnamefont {M.}~\bibnamefont {Vanderhaeghen}},\
  }\bibfield  {title} {\bibinfo {title} {{Leading-order QED radiative
  corrections to timelike Compton scattering on the proton}},\ }\href
  {https://doi.org/10.1103/PhysRevD.103.036009} {\bibfield  {journal} {\bibinfo
   {journal} {Phys. Rev. D}\ }\textbf {\bibinfo {volume} {103}},\ \bibinfo
  {pages} {036009} (\bibinfo {year} {2021})},\ \Eprint
  {https://arxiv.org/abs/2012.09565} {arXiv:2012.09565 [hep-ph]} \BibitemShut
  {NoStop}%
\bibitem [{\citenamefont {Goritschnig}\ \emph {et~al.}(2014)\citenamefont
  {Goritschnig}, \citenamefont {Pire},\ and\ \citenamefont
  {Wagner}}]{Goritschnig:2014eba}%
  \BibitemOpen
  \bibfield  {author} {\bibinfo {author} {\bibfnamefont {A.~T.}\ \bibnamefont
  {Goritschnig}}, \bibinfo {author} {\bibfnamefont {B.}~\bibnamefont {Pire}},\
  and\ \bibinfo {author} {\bibfnamefont {J.}~\bibnamefont {Wagner}},\
  }\bibfield  {title} {\bibinfo {title} {{Timelike Compton scattering with a
  linearly polarized photon beam}},\ }\href
  {https://doi.org/10.1103/PhysRevD.89.094031} {\bibfield  {journal} {\bibinfo
  {journal} {Phys. Rev. D}\ }\textbf {\bibinfo {volume} {89}},\ \bibinfo
  {pages} {094031} (\bibinfo {year} {2014})},\ \Eprint
  {https://arxiv.org/abs/1404.0713} {arXiv:1404.0713 [hep-ph]} \BibitemShut
  {NoStop}%
\bibitem [{\citenamefont {Bo{\"e}r}\ \emph {et~al.}(2015)\citenamefont
  {Bo{\"e}r}, \citenamefont {Guidal},\ and\ \citenamefont
  {Vanderhaeghen}}]{Boer:2015fwa}%
  \BibitemOpen
  \bibfield  {author} {\bibinfo {author} {\bibfnamefont {M.}~\bibnamefont
  {Bo{\"e}r}}, \bibinfo {author} {\bibfnamefont {M.}~\bibnamefont {Guidal}},\
  and\ \bibinfo {author} {\bibfnamefont {M.}~\bibnamefont {Vanderhaeghen}},\
  }\bibfield  {title} {\bibinfo {title} {{Timelike Compton scattering off the
  proton and generalized parton distributions}},\ }\href
  {https://doi.org/10.1140/epja/i2015-15103-3} {\bibfield  {journal} {\bibinfo
  {journal} {Eur. Phys. J. A}\ }\textbf {\bibinfo {volume} {51}},\ \bibinfo
  {pages} {103} (\bibinfo {year} {2015})}\BibitemShut {NoStop}%
\bibitem [{\citenamefont {Pire}\ \emph {et~al.}(2011)\citenamefont {Pire},
  \citenamefont {Szymanowski},\ and\ \citenamefont {Wagner}}]{Pire:2011st}%
  \BibitemOpen
  \bibfield  {author} {\bibinfo {author} {\bibfnamefont {B.}~\bibnamefont
  {Pire}}, \bibinfo {author} {\bibfnamefont {L.}~\bibnamefont {Szymanowski}},\
  and\ \bibinfo {author} {\bibfnamefont {J.}~\bibnamefont {Wagner}},\
  }\bibfield  {title} {\bibinfo {title} {{NLO corrections to timelike,
  spacelike and double deeply virtual Compton scattering}},\ }\href
  {https://doi.org/10.1103/PhysRevD.83.034009} {\bibfield  {journal} {\bibinfo
  {journal} {Phys. Rev. D}\ }\textbf {\bibinfo {volume} {83}},\ \bibinfo
  {pages} {034009} (\bibinfo {year} {2011})},\ \Eprint
  {https://arxiv.org/abs/1101.0555} {arXiv:1101.0555 [hep-ph]} \BibitemShut
  {NoStop}%
\bibitem [{\citenamefont {Diehl}(2001)}]{Diehl:2001pm}%
  \BibitemOpen
  \bibfield  {author} {\bibinfo {author} {\bibfnamefont {M.}~\bibnamefont
  {Diehl}},\ }\bibfield  {title} {\bibinfo {title} {{Generalized parton
  distributions with helicity flip}},\ }\href
  {https://doi.org/10.1007/s100520100635} {\bibfield  {journal} {\bibinfo
  {journal} {Eur. Phys. J. C}\ }\textbf {\bibinfo {volume} {19}},\ \bibinfo
  {pages} {485} (\bibinfo {year} {2001})},\ \Eprint
  {https://arxiv.org/abs/hep-ph/0101335} {arXiv:hep-ph/0101335} \BibitemShut
  {NoStop}%
\bibitem [{\citenamefont {Belitsky}\ and\ \citenamefont
  {Mueller}(2000)}]{Belitsky:2000jk}%
  \BibitemOpen
  \bibfield  {author} {\bibinfo {author} {\bibfnamefont {A.~V.}\ \bibnamefont
  {Belitsky}}\ and\ \bibinfo {author} {\bibfnamefont {D.}~\bibnamefont
  {Mueller}},\ }\bibfield  {title} {\bibinfo {title} {{Off forward
  gluonometry}},\ }\href {https://doi.org/10.1016/S0370-2693(00)00773-5}
  {\bibfield  {journal} {\bibinfo  {journal} {Phys. Lett. B}\ }\textbf
  {\bibinfo {volume} {486}},\ \bibinfo {pages} {369} (\bibinfo {year}
  {2000})},\ \Eprint {https://arxiv.org/abs/hep-ph/0005028}
  {arXiv:hep-ph/0005028} \BibitemShut {NoStop}%
\bibitem [{\citenamefont {Hoodbhoy}\ and\ \citenamefont
  {Ji}(1998)}]{Hoodbhoy:1998vm}%
  \BibitemOpen
  \bibfield  {author} {\bibinfo {author} {\bibfnamefont {P.}~\bibnamefont
  {Hoodbhoy}}\ and\ \bibinfo {author} {\bibfnamefont {X.-D.}\ \bibnamefont
  {Ji}},\ }\bibfield  {title} {\bibinfo {title} {{Helicity flip off forward
  parton distributions of the nucleon}},\ }\href
  {https://doi.org/10.1103/PhysRevD.58.054006} {\bibfield  {journal} {\bibinfo
  {journal} {Phys. Rev. D}\ }\textbf {\bibinfo {volume} {58}},\ \bibinfo
  {pages} {054006} (\bibinfo {year} {1998})},\ \Eprint
  {https://arxiv.org/abs/hep-ph/9801369} {arXiv:hep-ph/9801369} \BibitemShut
  {NoStop}%
\bibitem [{\citenamefont {Diehl}\ \emph {et~al.}(1997)\citenamefont {Diehl},
  \citenamefont {Gousset}, \citenamefont {Pire},\ and\ \citenamefont
  {Ralston}}]{Diehl:1997bu}%
  \BibitemOpen
  \bibfield  {author} {\bibinfo {author} {\bibfnamefont {M.}~\bibnamefont
  {Diehl}}, \bibinfo {author} {\bibfnamefont {T.}~\bibnamefont {Gousset}},
  \bibinfo {author} {\bibfnamefont {B.}~\bibnamefont {Pire}},\ and\ \bibinfo
  {author} {\bibfnamefont {J.~P.}\ \bibnamefont {Ralston}},\ }\bibfield
  {title} {\bibinfo {title} {{Testing the handbag contribution to exclusive
  virtual Compton scattering}},\ }\href
  {https://doi.org/10.1016/S0370-2693(97)00959-3} {\bibfield  {journal}
  {\bibinfo  {journal} {Phys. Lett. B}\ }\textbf {\bibinfo {volume} {411}},\
  \bibinfo {pages} {193} (\bibinfo {year} {1997})},\ \Eprint
  {https://arxiv.org/abs/hep-ph/9706344} {arXiv:hep-ph/9706344} \BibitemShut
  {NoStop}%
\bibitem [{\citenamefont {Goloskokov}\ and\ \citenamefont
  {Kroll}(2007)}]{Goloskokov:2006hr}%
  \BibitemOpen
  \bibfield  {author} {\bibinfo {author} {\bibfnamefont {S.~V.}\ \bibnamefont
  {Goloskokov}}\ and\ \bibinfo {author} {\bibfnamefont {P.}~\bibnamefont
  {Kroll}},\ }\bibfield  {title} {\bibinfo {title} {{The Longitudinal
  cross-section of vector meson electroproduction}},\ }\href
  {https://doi.org/10.1140/epjc/s10052-007-0228-4} {\bibfield  {journal}
  {\bibinfo  {journal} {Eur. Phys. J. C}\ }\textbf {\bibinfo {volume} {50}},\
  \bibinfo {pages} {829} (\bibinfo {year} {2007})},\ \Eprint
  {https://arxiv.org/abs/hep-ph/0611290} {arXiv:hep-ph/0611290} \BibitemShut
  {NoStop}%
\bibitem [{\citenamefont {Goloskokov}\ and\ \citenamefont
  {Kroll}(2009)}]{Goloskokov:2008ib}%
  \BibitemOpen
  \bibfield  {author} {\bibinfo {author} {\bibfnamefont {S.~V.}\ \bibnamefont
  {Goloskokov}}\ and\ \bibinfo {author} {\bibfnamefont {P.}~\bibnamefont
  {Kroll}},\ }\bibfield  {title} {\bibinfo {title} {{The Target asymmetry in
  hard vector-meson electroproduction and parton angular momenta}},\ }\href
  {https://doi.org/10.1140/epjc/s10052-008-0833-x} {\bibfield  {journal}
  {\bibinfo  {journal} {Eur. Phys. J. C}\ }\textbf {\bibinfo {volume} {59}},\
  \bibinfo {pages} {809} (\bibinfo {year} {2009})},\ \Eprint
  {https://arxiv.org/abs/0809.4126} {arXiv:0809.4126 [hep-ph]} \BibitemShut
  {NoStop}%
\bibitem [{\citenamefont {Goloskokov}\ and\ \citenamefont
  {Kroll}(2010)}]{Goloskokov:2009ia}%
  \BibitemOpen
  \bibfield  {author} {\bibinfo {author} {\bibfnamefont {S.~V.}\ \bibnamefont
  {Goloskokov}}\ and\ \bibinfo {author} {\bibfnamefont {P.}~\bibnamefont
  {Kroll}},\ }\bibfield  {title} {\bibinfo {title} {{An Attempt to understand
  exclusive pi+ electroproduction}},\ }\href
  {https://doi.org/10.1140/epjc/s10052-009-1178-9} {\bibfield  {journal}
  {\bibinfo  {journal} {Eur. Phys. J. C}\ }\textbf {\bibinfo {volume} {65}},\
  \bibinfo {pages} {137} (\bibinfo {year} {2010})},\ \Eprint
  {https://arxiv.org/abs/0906.0460} {arXiv:0906.0460 [hep-ph]} \BibitemShut
  {NoStop}%
\bibitem [{\citenamefont {Wiringa}\ \emph {et~al.}(1995)\citenamefont
  {Wiringa}, \citenamefont {Stoks},\ and\ \citenamefont
  {Schiavilla}}]{Wiringa:1994wb}%
  \BibitemOpen
  \bibfield  {author} {\bibinfo {author} {\bibfnamefont {R.~B.}\ \bibnamefont
  {Wiringa}}, \bibinfo {author} {\bibfnamefont {V.~G.~J.}\ \bibnamefont
  {Stoks}},\ and\ \bibinfo {author} {\bibfnamefont {R.}~\bibnamefont
  {Schiavilla}},\ }\bibfield  {title} {\bibinfo {title} {{An Accurate
  nucleon-nucleon potential with charge independence breaking}},\ }\href
  {https://doi.org/10.1103/PhysRevC.51.38} {\bibfield  {journal} {\bibinfo
  {journal} {Phys. Rev. C}\ }\textbf {\bibinfo {volume} {51}},\ \bibinfo
  {pages} {38} (\bibinfo {year} {1995})},\ \Eprint
  {https://arxiv.org/abs/nucl-th/9408016} {arXiv:nucl-th/9408016} \BibitemShut
  {NoStop}%
\bibitem [{\citenamefont {Boussarie}\ and\ \citenamefont
  {Mehtar-Tani}(2024)}]{Boussarie:2023xun}%
  \BibitemOpen
  \bibfield  {author} {\bibinfo {author} {\bibfnamefont {R.}~\bibnamefont
  {Boussarie}}\ and\ \bibinfo {author} {\bibfnamefont {Y.}~\bibnamefont
  {Mehtar-Tani}},\ }\bibfield  {title} {\bibinfo {title} {{Low and moderate x
  gluon contribution to exclusive Compton scattering processes}},\ }\href
  {https://doi.org/10.1007/JHEP10(2024)056} {\bibfield  {journal} {\bibinfo
  {journal} {JHEP}\ }\textbf {\bibinfo {volume} {10}},\ \bibinfo {pages}
  {056}},\ \Eprint {https://arxiv.org/abs/2309.16576} {arXiv:2309.16576
  [hep-ph]} \BibitemShut {NoStop}%
\bibitem [{\citenamefont {Belitsky}\ \emph {et~al.}(2002)\citenamefont
  {Belitsky}, \citenamefont {Mueller},\ and\ \citenamefont
  {Kirchner}}]{Belitsky:2001ns}%
  \BibitemOpen
  \bibfield  {author} {\bibinfo {author} {\bibfnamefont {A.~V.}\ \bibnamefont
  {Belitsky}}, \bibinfo {author} {\bibfnamefont {D.}~\bibnamefont {Mueller}},\
  and\ \bibinfo {author} {\bibfnamefont {A.}~\bibnamefont {Kirchner}},\
  }\bibfield  {title} {\bibinfo {title} {{Theory of deeply virtual Compton
  scattering on the nucleon}},\ }\href
  {https://doi.org/10.1016/S0550-3213(02)00144-X} {\bibfield  {journal}
  {\bibinfo  {journal} {Nucl. Phys. B}\ }\textbf {\bibinfo {volume} {629}},\
  \bibinfo {pages} {323} (\bibinfo {year} {2002})},\ \Eprint
  {https://arxiv.org/abs/hep-ph/0112108} {arXiv:hep-ph/0112108} \BibitemShut
  {NoStop}%
\bibitem [{\citenamefont {Jaffe}\ and\ \citenamefont
  {Manohar}(1989)}]{Jaffe:1989xy}%
  \BibitemOpen
  \bibfield  {author} {\bibinfo {author} {\bibfnamefont {R.~L.}\ \bibnamefont
  {Jaffe}}\ and\ \bibinfo {author} {\bibfnamefont {A.}~\bibnamefont
  {Manohar}},\ }\bibfield  {title} {\bibinfo {title} {{NUCLEAR GLUONOMETRY}},\
  }\href {https://doi.org/10.1016/0370-2693(89)90242-6} {\bibfield  {journal}
  {\bibinfo  {journal} {Phys. Lett. B}\ }\textbf {\bibinfo {volume} {223}},\
  \bibinfo {pages} {218} (\bibinfo {year} {1989})}\BibitemShut {NoStop}%
\bibitem [{\citenamefont {Nzar}\ and\ \citenamefont
  {Hoodbhoy}(1992)}]{Nzar:1992ax}%
  \BibitemOpen
  \bibfield  {author} {\bibinfo {author} {\bibfnamefont {M.}~\bibnamefont
  {Nzar}}\ and\ \bibinfo {author} {\bibfnamefont {P.}~\bibnamefont
  {Hoodbhoy}},\ }\bibfield  {title} {\bibinfo {title} {{Estimation of the
  double helicity flip deuteron structure function}},\ }\href
  {https://doi.org/10.1103/PhysRevD.45.2264} {\bibfield  {journal} {\bibinfo
  {journal} {Phys. Rev. D}\ }\textbf {\bibinfo {volume} {45}},\ \bibinfo
  {pages} {2264} (\bibinfo {year} {1992})}\BibitemShut {NoStop}%
\bibitem [{\citenamefont {Kumano}\ and\ \citenamefont
  {Song}(2020{\natexlab{a}})}]{Kumano:2020gfk}%
  \BibitemOpen
  \bibfield  {author} {\bibinfo {author} {\bibfnamefont {S.}~\bibnamefont
  {Kumano}}\ and\ \bibinfo {author} {\bibfnamefont {Q.-T.}\ \bibnamefont
  {Song}},\ }\bibfield  {title} {\bibinfo {title} {{Deuteron polarizations in
  the proton-deuteron Drell-Yan process for finding the gluon transversity}},\
  }\href {https://doi.org/10.1103/PhysRevD.101.094013} {\bibfield  {journal}
  {\bibinfo  {journal} {Phys. Rev. D}\ }\textbf {\bibinfo {volume} {101}},\
  \bibinfo {pages} {094013} (\bibinfo {year} {2020}{\natexlab{a}})},\ \Eprint
  {https://arxiv.org/abs/2003.06623} {arXiv:2003.06623 [hep-ph]} \BibitemShut
  {NoStop}%
\bibitem [{\citenamefont {Kumano}\ and\ \citenamefont
  {Song}(2020{\natexlab{b}})}]{Kumano:2019igu}%
  \BibitemOpen
  \bibfield  {author} {\bibinfo {author} {\bibfnamefont {S.}~\bibnamefont
  {Kumano}}\ and\ \bibinfo {author} {\bibfnamefont {Q.-T.}\ \bibnamefont
  {Song}},\ }\bibfield  {title} {\bibinfo {title} {{Gluon transversity in
  polarized proton-deuteron Drell-Yan process}},\ }\href
  {https://doi.org/10.1103/PhysRevD.101.054011} {\bibfield  {journal} {\bibinfo
   {journal} {Phys. Rev. D}\ }\textbf {\bibinfo {volume} {101}},\ \bibinfo
  {pages} {054011} (\bibinfo {year} {2020}{\natexlab{b}})},\ \Eprint
  {https://arxiv.org/abs/1910.12523} {arXiv:1910.12523 [hep-ph]} \BibitemShut
  {NoStop}%
\bibitem [{\citenamefont {Winter}\ \emph {et~al.}(2017)\citenamefont {Winter},
  \citenamefont {Detmold}, \citenamefont {Gambhir}, \citenamefont {Orginos},
  \citenamefont {Savage}, \citenamefont {Shanahan},\ and\ \citenamefont
  {Wagman}}]{Winter:2017bfs}%
  \BibitemOpen
  \bibfield  {author} {\bibinfo {author} {\bibfnamefont {F.}~\bibnamefont
  {Winter}}, \bibinfo {author} {\bibfnamefont {W.}~\bibnamefont {Detmold}},
  \bibinfo {author} {\bibfnamefont {A.~S.}\ \bibnamefont {Gambhir}}, \bibinfo
  {author} {\bibfnamefont {K.}~\bibnamefont {Orginos}}, \bibinfo {author}
  {\bibfnamefont {M.~J.}\ \bibnamefont {Savage}}, \bibinfo {author}
  {\bibfnamefont {P.~E.}\ \bibnamefont {Shanahan}},\ and\ \bibinfo {author}
  {\bibfnamefont {M.~L.}\ \bibnamefont {Wagman}},\ }\bibfield  {title}
  {\bibinfo {title} {{First lattice QCD study of the gluonic structure of light
  nuclei}},\ }\href {https://doi.org/10.1103/PhysRevD.96.094512} {\bibfield
  {journal} {\bibinfo  {journal} {Phys. Rev. D}\ }\textbf {\bibinfo {volume}
  {96}},\ \bibinfo {pages} {094512} (\bibinfo {year} {2017})},\ \Eprint
  {https://arxiv.org/abs/1709.00395} {arXiv:1709.00395 [hep-lat]} \BibitemShut
  {NoStop}%
\bibitem [{\citenamefont {Maxwell}\ \emph {et~al.}(2018)\citenamefont {Maxwell}
  \emph {et~al.}}]{Maxwell:2018gci}%
  \BibitemOpen
  \bibfield  {author} {\bibinfo {author} {\bibfnamefont {J.}~\bibnamefont
  {Maxwell}} \emph {et~al.},\ }\bibfield  {title} {\bibinfo {title} {{Search
  for Exotic Gluonic States in the Nucleus, A Letter of Intent to Jefferson Lab
  PAC 44}},\ }\href@noop {} {\  (\bibinfo {year} {2018})},\ \Eprint
  {https://arxiv.org/abs/1803.11206} {arXiv:1803.11206 [nucl-ex]} \BibitemShut
  {NoStop}%
\end{thebibliography}%


\end{document}